\documentclass[lettersize,journal]{IEEEtran}
\usepackage{amsmath,amsfonts}
\usepackage{algorithmic}
\usepackage{algorithm}
\usepackage{array}
\usepackage[caption=false,font=normalsize,labelfont=sf,textfont=sf]{subfig}
\usepackage{textcomp}
\usepackage{stfloats}
\usepackage{url}
\usepackage{verbatim}
\usepackage{graphicx}
\usepackage{cite}
\begin{document}

\title{Stereo Audio Rendering for Personal Sound Zones Using a Binaural Spatially Adaptive Neural Network (BSANN)}

\author{
Hao~Jiang, Edgar~Choueiri%
\thanks{
This work was supported by a research grant from Sound United Corporation.
(Corresponding author: Hao Jiang.)
}%
\thanks{
The authors are with the 3D Audio and Applied Acoustics (3D3A) Laboratory,
Princeton University, Princeton, NJ 08544 USA
(e-mail: hj3737@princeton.edu; choueiri@princeton.edu).
}%
}

\markboth{IEEE Transactions on Audio, Speech, and Language Processing,~Vol.~XX,~XXXX}%
{Jiang \MakeLowercase{\textit{et al.}}: Stereo Audio Rendering for Personal Sound Zones Using a Binaural Spatially Adaptive Neural Network (BSANN)}

\maketitle

\begin{abstract}
A binaural rendering framework for personal sound zones (PSZs) is proposed to enable multiple head-tracked listeners to receive fully independent stereo audio programs. Current PSZ systems typically rely on monophonic rendering and therefore cannot control the left and right ears separately, which limits the quality and accuracy of spatial imaging. The proposed method employs a Binaural Spatially Adaptive Neural Network (BSANN) to generate ear-optimized loudspeaker filters that reconstruct the desired acoustic field at each ear of multiple listeners. The framework integrates anechoically measured loudspeaker frequency responses, analytically modeled transducer directivity, and rigid-sphere head-related transfer functions (HRTFs) to enhance acoustic accuracy and spatial rendering fidelity. An explicit active crosstalk cancellation (XTC) stage further improves three-dimensional spatial perception.  Experiments show significant gains in measured objective performance metrics, including inter-zone isolation (IZI), inter-program isolation (IPI), and crosstalk cancellation (XTC), with log-frequency–weighted values of 10.23/10.03~dB (IZI), 11.11/9.16~dB (IPI), and 10.55/11.13~dB (XTC), respectively, over 100--20{,}000~Hz. The combined use of ear-wise control, accurate acoustic modeling, and integrated active XTC produces a unified rendering method that delivers greater isolation performance, increased robustness to room asymmetry, and more faithful spatial reproduction in real acoustic environments.
\end{abstract}



\section{Introduction}

\IEEEPARstart{P}{ersonal} sound zones (PSZs)~\cite{druyvesteyn1997personal,betlehem2015personal} aim to deliver independent audio programs to multiple listeners within a shared acoustic space while minimizing interference between listeners. Applications include home entertainment systems~\cite{jacobsen2023living}, automotive audio~\cite{cheer2013car,vindrola2021car}, healthcare and public environments~\cite{skov2023hospital}, and large-scale outdoor installations~\cite{heuchel2020large}. A typical PSZ system defines a bright zone (BZ), in which the target signal is reproduced with high fidelity, and a dark zone (DZ), in which unwanted signals are suppressed. Classical filter-design methods such as acoustic contrast control (ACC)~\cite{choi2002brightzone,galvez2015timedomain}, pressure matching (PM)~\cite{poletti2008multizone,chang2012doublelayer}, amplitude matching (AM)~\cite{abe2023amplitude}, and variable span trade-off filtering (VAST)~\cite{lee2020signaladaptive,brunnstrom2022vast} have been extensively developed and provide effective control under static listening conditions.

Despite significant progress, traditional PSZ methods still face fundamental limitations. Because they rely on a fixed set of measured acoustic transfer functions (ATFs), their performance degrades whenever environmental variations or listener head movements introduce ATF mismatch. Room asymmetry, loudspeaker variability, measurement noise, and insufficient spatial sampling have all been shown to reduce achievable isolation performance~\cite{elliott2012robustness,coleman2014acousticcontrast,zhu2017robustacc,moller2019tfnoise}. In addition, closed-form filter-design formulations, in contrast to neural network methods, typically require repeated matrix inversion across frequencies and listener positions, leading to substantial computational cost~\cite{lee2020signaladaptive}. Moreover, recent studies have shown that isolation performance depends strongly on the full positional configuration of the listeners’ heads, including both the location and orientation of each head (its pose, i.e., its six-degree-of-freedom position and orientation)~\cite{qiao2023aes_brtf,galvez2019dynamic}. The resulting pose space is exceedingly high-dimensional, making it infeasible for closed-form methods to precompute or densely sample pose-specific filters. This makes closed-form filter computation impractical for head-tracked PSZ, since every new head pose would require generating a new set of filters. Consequently, approaches that rely on pre-designed or interpolated filters do not naturally scale to dynamic, head-tracked rendering or to spatially accurate stereo and binaural reproduction.

Deep-learning-based approaches have recently emerged as promising alternatives for sound field control~\cite{pepe2022neuralpsz,alessandri2021deep}. The Spatially Adaptive Neural Network (SANN) introduced in \cite{qiao2025sann} demonstrated that PSZ filters can be generated directly from listener coordinates, eliminating per-position matrix inversion, and improving robustness across simulated and real environments. However, as a monophonic formulation, the SANN does not preserve stereo imagery or interaural spatial cues, both of which are essential for realistic spatial audio reproduction. In addition, its reliance on point-source ATFs simulations limits physical accuracy and reduces isolation performance when deployed with real loudspeaker arrays; incorporating measured ATFs during fine-tuning can improve realism, but this calibration requires extensive in-situ measurements and therefore limits practical deployment.

To overcome these limitations, we introduce the Binaural Spatially Adaptive Neural Network (BSANN), a framework for binaural audio rendering in PSZs. The BSANN extends the original monophonic architecture by independently optimizing the acoustic field at each ear of multiple head-tracked listeners. This ear-wise control mitigates the sensitivity of PSZ systems to room asymmetry by explicitly regulating all listeners' ears, thereby stabilizing isolation performance in practical acoustic environments. To improve fidelity to real loudspeaker--listener systems, the framework integrates anechoically measured loudspeaker frequency responses, analytically modeled transducer directivity~\cite{morse1968theoretical,Stepanishen1971Transient}, and rigid-sphere HRTFs~\cite{kuhn1977spherical,duda1998spherical,qiao2023aes_brtf}. These components collectively yield more accurate training data and stronger generalization to real acoustic conditions. Furthermore, although the use of measured loudspeaker frequency responses ties the method to a specific loudspeaker array design, it does not impose any listening-environment constraints beyond those already assumed during network training. Because the measurements are anechoic, they generalize well across different rooms.

The framework further incorporates an explicit active crosstalk cancellation (XTC) stage~\cite{ma2019superdirective,choueiri2018binaural,kabzinski2019adaptive} to enable loudspeaker-based binaural rendering with improved spatial fidelity~\cite{galvez2019dynamic,lindfors2022equalization}. This integration naturally supports head-tracked binaural rendering and complements the ear-wise PSZ objectives, forming a unified stereo and binaural architecture that enables more faithful transmission of spatial cues, including interaural time differences (ITDs) and interaural level differences (ILDs).

The contributions of this work are threefold:
\begin{enumerate}
    \item We propose the BSANN, which generalizes the monophonic SANN to an ear-wise stereo architecture. By explicitly controlling all listeners' ears, the BSANN improves robustness to room asymmetry while enabling stereo and binaural PSZ rendering with higher and more consistent isolation performance.
    \item We enhance the acoustic realism of the training data by combining simulated ATFs with anechoically measured loudspeaker responses, analytically modeled transducer directivity, and rigid-sphere HRTFs. This physically informed modeling improves the physical accuracy of the learned filters and yields higher and more consistent inter-zone isolation (IZI) and inter-program isolation (IPI), as well as better crosstalk cancellation under practical acoustic conditions.
    \item We integrate an explicit active XTC stage to achieve accurate spatial rendering and strong crosstalk cancellation, enabling high-fidelity three-dimensional imaging for head-tracked listeners.
\end{enumerate}

The rest of the paper is organized as follows. 
In Sec.~\ref{sec:method}, we present the proposed BSANN framework, including the ear-wise formulation, network architecture, physically informed acoustic modeling, and the integration of the active XTC stage. 
In Sec.~\ref{sec:experiment}, we describe the loudspeaker configuration, listener geometry, training dataset, and the objective evaluation metrics. 
In Sec.~\ref{sec:results}, we report the measured objective evaluation results and compare the proposed method with existing approaches. 
Finally, Sec.~\ref{sec:conclusion} summarizes the findings and discusses the implications of the stereo and binaural PSZ framework, and provides brief additional methodological remarks, with detailed analysis left for future work.


\section{Methodology}
\label{sec:method}

\subsection{Binaural Spatially Adaptive Neural Network (BSANN)}
\label{sec:BSANN}

\begin{figure*}[!t]
    \centering
    \includegraphics[width=\textwidth]{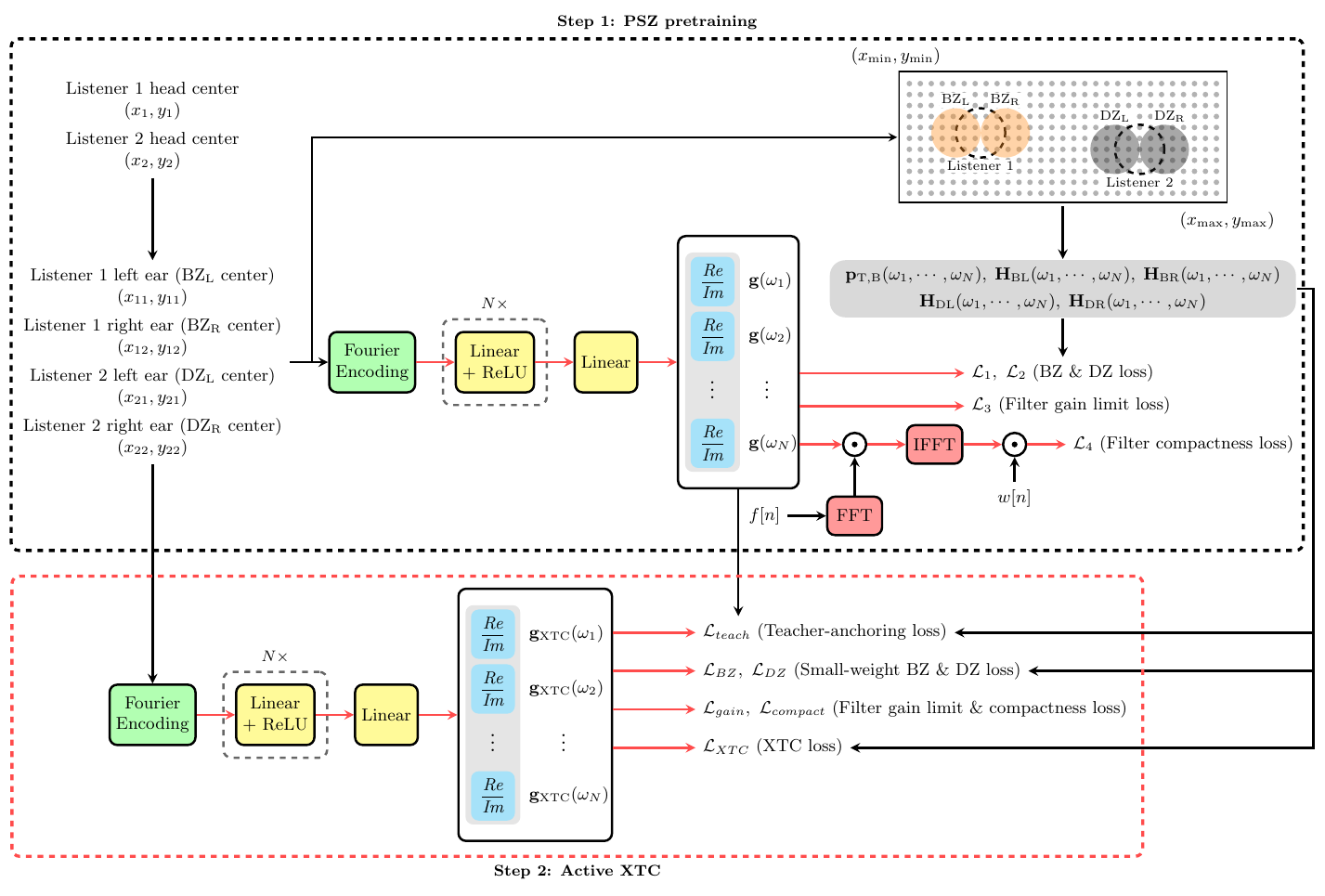}
    \caption{Overview of the proposed two-stage BSANN framework.
    Step~1 performs PSZ pretraining to obtain the frequency-domain loudspeaker
    filters $\mathbf{g}(\omega_n)$ under a loss function composed of the 
    bright-zone and dark-zone objectives together with the gain and compactness 
    constraints.
    Step~2 refines the filters to $\mathbf{g}_{\mathrm{XTC}}(\omega_n)$ under a loss function dominated by the active XTC term and regularized by teacher-anchoring, small-weight BZ/DZ protector terms, and the gain and compactness constraints.}
    \label{fig:BSANN}
\end{figure*}

The proposed BSANN framework extends the monophonic SANN model introduced in~\cite{qiao2025sann} to enable ear-wise control for multiple head-tracked listeners. In this work, we focus on the two-listener case, while noting that the method can be readily extended to more. Let $X_{\ell}(\omega)$ be the driving signal for loudspeaker~$\ell$, and let 
$H_{e,m,\ell}(\omega)$ denote the acoustic transfer function (ATF) from loudspeaker~$\ell$ to control point $m=1,\dots,M$ surrounding ear 
$e\in\{\mathrm{L}_1,\mathrm{R}_1,\mathrm{L}_2,\mathrm{R}_2\}$.
The reproduced pressure at control point $m$ of ear $e$ is
\begin{equation}
    Y_{e,m}(\omega)
    =
    \sum_{\ell=1}^{L}
    H_{e,m,\ell}(\omega)\,X_{\ell}(\omega).
\end{equation}
For stereo PSZ rendering, the left and right program channels for each listener 
are treated as independent inputs, and the BSANN regulates how these channels 
contribute to all control points of the ears in accordance with the 
sound-zone objectives.

The BSANN predicts the complex-valued loudspeaker filter bank
\begin{equation}
    \mathbf{g}(\omega)= [g_1(\omega),\dots,g_L(\omega)]^{\mathsf{T}},
\end{equation}
conditioned on the positions of the two listeners. Let $\mathbf{S}(\omega)$ denote the four input program channels corresponding to
the left and right program signals intended for Listener~1 and Listener~2:
\[
\mathbf{S}(\omega)=
\begin{bmatrix}
S_{1,\mathrm{L}}(\omega)\\
S_{1,\mathrm{R}}(\omega)\\
S_{2,\mathrm{L}}(\omega)\\
S_{2,\mathrm{R}}(\omega)
\end{bmatrix},
\qquad
p\in\{1,2,3,4\}.
\]
For each loudspeaker $\ell$, the BSANN predicts complex weights
$g_{\ell,p}(\omega)$ that determine the contribution of each program channel.  
Thus,
\begin{equation}
    X_{\ell}(\omega)
    = \sum_{p=1}^{4} g_{\ell,p}(\omega)\, S_{p}(\omega),
    \label{eq:xl_sum}
\end{equation}
or, equivalently,
\begin{equation}
    X_{\ell}(\omega)
    = \mathbf{g}_{\ell}^{\mathsf{T}}(\omega)\,\mathbf{S}(\omega),
\end{equation}
where $\mathbf{g}_{\ell}(\omega)=[g_{\ell,1}(\omega),\dots,g_{\ell,4}(\omega)]^{\mathsf{T}}$.

The mapping, illustrated in Fig.~\ref{fig:BSANN}, consists of Fourier positional 
encoding followed by an $N$-layer shared multi-layer perceptron (MLP)~\cite{tancik2020fourier,goodfellow2016deep} and a final linear layer that 
outputs the real and imaginary components of $\mathbf{g}(\omega_n)$ at all discrete frequency bins. 
The nonlinear transformation is represented as
\begin{equation}
    \mathbf{g}(\omega_n)=f_{\theta}(\mathbf{s}),
\end{equation}
where $\mathbf{s}$ denotes the listener head positions (with ear positions 
obtained geometrically) and $\theta$ denotes the trainable network parameters.

To support ear-wise stereo reproduction, the BSANN adopts a loss function 
that generalizes the monophonic SANN objective to the four-ear stereo setting. 
For discrete frequency bins $\{\omega_n\}_{n=1}^{N}$, the loss terms are defined as follows.

\noindent\textit{1) Bright-zone accuracy:}
Let
\begin{equation}
\tilde{\mathbf{p}}_{\mathrm{T},B}(\omega_n)=
\begin{bmatrix}
\mathbf{p}_{\mathrm{T,BL}}(\omega_n)\\[2pt]
\mathbf{p}_{\mathrm{T,BR}}(\omega_n)
\end{bmatrix},
\;
\tilde{\mathbf{H}}_{\mathrm{B}}(\omega_n)=
\begin{bmatrix}
\mathbf{H}_{\mathrm{BL}}(\omega_n)\\[2pt]
\mathbf{H}_{\mathrm{BR}}(\omega_n)
\end{bmatrix},
\end{equation}
where $\mathbf{p}_{\mathrm{T,BL}}$ and $\mathbf{p}_{\mathrm{T,BR}}$ denote
the target sound pressure for the left and right bright-zone ears, and 
$\mathbf{H}_{\mathrm{BL}}$, $\mathbf{H}_{\mathrm{BR}}$ are the corresponding
ear-specific ATF submatrices.

The bright-zone loss is defined as
\begin{equation}
\mathcal{L}_1
=
\frac{1}{N M_{\mathrm{B}}}
\sum_{n=1}^{N}
\Big\|\,
\big|\tilde{\mathbf{p}}_{\mathrm{T},B}(\omega_n)\big|
-
\big|\tilde{\mathbf{H}}_{\mathrm{B}}(\omega_n)\mathbf{g}(\omega_n)\big|
\Big\|_2^{2}.
\end{equation}
Here $M_{\mathrm{B}}$ denotes the number of bright-zone control points
per ear (assumed identical for the left and right ears), and $N$ is the
number of discrete frequency bins over which the loss is averaged.

\noindent\textit{2) Dark-zone suppression:}
Define
\begin{equation}
\tilde{\mathbf{H}}_{\mathrm{D}}(\omega_n)=
\begin{bmatrix}
\mathbf{H}_{\mathrm{DL}}(\omega_n)\\
\mathbf{H}_{\mathrm{DR}}(\omega_n)
\end{bmatrix},
\end{equation}
corresponding to the dark-zone ears of the inactive listener.
The dark-zone loss is
\begin{equation}
\mathcal{L}_2
=
\frac{1}{N M_{\mathrm{D}}}
\sum_{n=1}^{N}
\Big\|
\tilde{\mathbf{H}}_{\mathrm{D}}(\omega_n)\mathbf{g}(\omega_n)
\Big\|_2^{2},
\end{equation}
where the target pressure for the DZ is assumed to be zero.

\noindent\textit{3) Gain limiting:}
To avoid excessively large filter magnitudes, a gain constraint is applied:
\begin{equation}
\mathcal{L}_3
=
\frac{1}{NL}
\sum_{n=1}^{N}
\Big\|
\max\!\big(\mathbf{0},\,|\mathbf{g}(\omega_n)| - g_{\max}\mathbf{1}\big)
\Big\|_2^{2},
\end{equation}
where $|\mathbf{g}(\omega_n)|$ is the elementwise magnitude, $g_{\max}$ is the gain limit, $\mathbf{0}$ and $\mathbf{1}$ are all-zero and all-one vectors of length $L$, and $L$ is the number of loudspeakers.

\noindent\textit{4) Time-domain compactness:}
To discourage long and oscillatory filters in the time domain, a tapered window is applied after inverse FFT to penalize late energy:
\begin{equation}
\mathcal{L}_4
=
\frac{1}{\hat{N} L}
\sum_{\ell=1}^{L}
\big\|
w[n]\cdot\big(f[n]*\hat{g}_{\ell}[n]\big)
\big\|_2^{2},
\end{equation}
where $f[n]$ is a bandpass weighting filter and $\hat{N}$ is the time-domain filter length.

The combined objective is
\begin{equation}
\mathcal{L}
=
\alpha\,\mathcal{L}_1
+
(1-\alpha)\,\mathcal{L}_2
+
\beta\,\mathcal{L}_3
+
\gamma\,\mathcal{L}_4,
\label{eq:lossPSZ}
\end{equation}
where $\alpha$, $\beta$, and $\gamma$ are weighting parameters that
balance the contributions of the four terms.

\subsection{Physically Informed ATF Construction}
\label{sec:physmodel}

\begin{figure*}[!t]
    \centering
    \includegraphics[width=\textwidth]{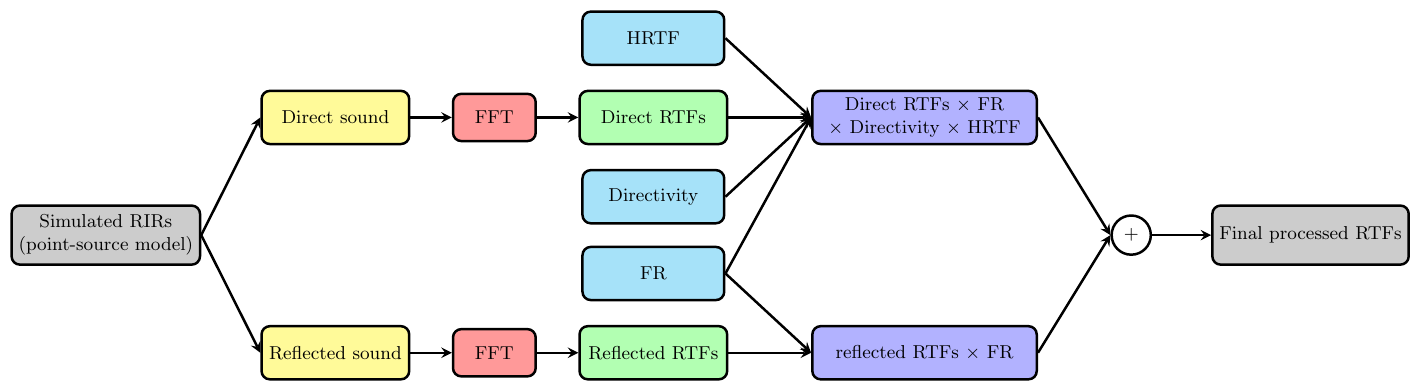}
    \caption{Processing pipeline used to construct the physically informed
    acoustic transfer functions (ATFs) for BSANN training. 
    Simulated room impulse responses (RIRs) are decomposed into direct and 
    reflected components, transformed into frequency-domain room transfer 
    functions (RTFs), and combined with measured anechoic responses, 
    analytic piston directivity, and rigid-sphere HRTFs to obtain the ATFs 
    used in the BSANN loss functions.}
    \label{fig:atf_flowchart}
\end{figure*}

The ATFs used for BSANN training are constructed from three physically
informed components: anechoically measured loudspeaker frequency responses, an analytic piston directivity model that approximates the physical transducer directivity, and rigid-sphere HRTFs. Compared to ideal point-source ATFs, this physically informed modeling captures more realistic frequency–angle behavior. The analytic piston directivity and rigid-sphere HRTFs are purely mathematical and therefore introduce no dependence on any particular array or listening environment. The loudspeaker frequency responses, which are the only measured component, are tied to a specific loudspeaker array design; however, because they are obtained anechoically, they do not impose any listening-environment constraints beyond those already assumed during network training. Figure~\ref{fig:atf_flowchart} summarizes the overall processing pipeline, which consists of three major components:

\noindent\textit{1) Anechoic loudspeaker frequency responses:}
Let $a_{\ell}[n]$ denote the measured anechoic impulse response of loudspeaker $\ell$. The corresponding frequency response is
\begin{equation}
A_{\ell}(\omega) = \mathrm{FFT}\{a_{\ell}[n]\},
\end{equation}
which captures the loudspeaker-specific spectral coloration. These responses ensure that the BSANN is trained with spectra that are representative of the deployed hardware, rather than with ideal flat responses.

\noindent\textit{2) Analytic piston directivity:}
The directional radiation of each loudspeaker is modeled analytically using the circular-piston model in a rigid baffle. For a piston of radius $a$, the directivity factor as a function of angular frequency $\omega$ and observation angle
$\theta$ is given by
\begin{equation}
D_{\ell}(\omega,\theta)
=
\frac{2 J_{1}(k a \sin\theta)}{k a \sin\theta},
\qquad
k = \frac{\omega}{c},
\end{equation}
where $J_{1}(\cdot)$ is the first-order Bessel function and $c$ is the speed of sound. This model reproduces the frequency-dependent beam narrowing of finite-size loudspeakers without requiring directivity measurements.

\noindent\textit{3) Rigid-sphere HRTFs.}
The ear-adjacent control points are modeled using a normalized rigid-sphere
scattering formulation. Let $\mathbf{r}_{\mathrm{src}}$ denote the direction
from the loudspeaker to the head center (i.e., each loudspeaker is treated
as a point source in the rigid-sphere model), and let
$\mathbf{r}_{e,m}$ denote the position of control point $m$ around ear $e$
relative to the head center, with radii
$r_{\mathrm{src}} = \|\mathbf{r}_{\mathrm{src}}\|$,
$r_{e,m} = \|\mathbf{r}_{e,m}\|$, and angular separation
$\gamma_{e,m} = \angle(\mathbf{r}_{\mathrm{src}},\mathbf{r}_{e,m})$.
For wavenumber $k = \omega/c$, define the scattering kernel
\begin{equation}
\label{eq:S_l}
\begin{aligned}
S_n(k)
=\;& j_n(k r_<)\,h^{(1)}_n(k r_>) \\
&- \alpha_n(ka)\,
   h^{(1)}_n(k r_{\mathrm{src}})\,
   h^{(1)}_n(k r_{e,m}),
\end{aligned}
\end{equation}
where $r_< = \min(r_{\mathrm{src}},r_{e,m})$,
$r_> = \max(r_{\mathrm{src}},r_{e,m})$,
$j_n(\cdot)$ and $h^{(1)}_n(\cdot)$ are spherical Bessel and Hankel
functions, and
$\alpha_n(ka)=j'_n(ka)/h^{(1)\,'}_n(ka)$ denotes the rigid-sphere
scattering coefficient of order $n$.

The normalized rigid-sphere HRTF from the loudspeaker direction to control
point $m$ around ear $e$ is
\begin{equation}
\label{eq:rigidsphere_hrtf}
H_{\mathrm{HRTF},e,m}(\omega)
=
\frac{j k}{G_{\mathrm{ff},e,m}(\omega)}
\sum_{n=0}^{N_{\mathrm{max}}}(2n+1)\,
S_n(k)\,
P_n(\cos\gamma_{e,m}),
\end{equation}
where $P_n(\cdot)$ is the Legendre polynomial and
\begin{equation}
G_{\mathrm{ff},e,m}(\omega)
=
\frac{
e^{-jk\|\mathbf{r}_{\mathrm{src}}-\mathbf{r}_{e,m}\|}
}{
\|\mathbf{r}_{\mathrm{src}}-\mathbf{r}_{e,m}\|
}
\end{equation}
is the free-field Green’s function used for normalization.

\noindent\textit{4) Final ATF assembly.}
For each loudspeaker–ear–control-point triplet $(\ell,e,m)$, the simulated
point-source room impulse response is decomposed into a direct component
$h^{\mathrm{dir}}_{e,m,\ell}[n]$ and a reflected component
$h^{\mathrm{refl}}_{e,m,\ell}[n]$, as illustrated in
Fig.~\ref{fig:atf_flowchart}.  
The corresponding frequency-domain transfer functions are
\begin{equation}
\label{eq:dir_refl}
\begin{aligned}
H^{\mathrm{dir}}_{e,m,\ell}(\omega)
&= \mathrm{FFT}\{h^{\mathrm{dir}}_{e,m,\ell}[n]\}, \\
H^{\mathrm{refl}}_{e,m,\ell}(\omega)
&= \mathrm{FFT}\{h^{\mathrm{refl}}_{e,m,\ell}[n]\}.
\end{aligned}
\end{equation}

The physically informed ATF used for BSANN training is then obtained by applying loudspeaker frequency responses, analytic piston directivity, and rigid-sphere
HRTFs to the direct path, and loudspeaker frequency responses to the reflected path:
\begin{equation}
\label{eq:atf_flow}
\begin{aligned}
\tilde{H}_{e,m,\ell}(\omega)
&=
H^{\mathrm{dir}}_{e,m,\ell}(\omega)\,
A_{\ell}(\omega)\,
D_{\ell}(\omega,\theta_{e,m,\ell})\,
H_{\mathrm{HRTF},e,m}(\omega) \\
&\quad+
H^{\mathrm{refl}}_{e,m,\ell}(\omega)\,
A_{\ell}(\omega),
\end{aligned}
\end{equation}
where $A_{\ell}(\omega)$ is the anechoically measured loudspeaker frequency response,
$D_{\ell}(\omega,\theta_{e,m,\ell})$ is the analytic piston directivity
evaluated at the angle $\theta_{e,m,\ell}$ from loudspeaker~$\ell$ to
control point $m$ of ear $e$, and $H_{\mathrm{HRTF},e,m}(\omega)$ is the
normalized rigid-sphere HRTF defined above. The direct term captures directional effects and ear-adjacent filtering, whereas the reflected term contributes nondirectional late energy.
Evaluating~\eqref{eq:atf_flow} over all control points
$m=1,\dots,M$ for each ear $e$ and all loudspeakers $\ell$
yields the ear-wise ATF submatrices
$\tilde{\mathbf{H}}_{\mathrm{BL}}$,
$\tilde{\mathbf{H}}_{\mathrm{BR}}$,
$\tilde{\mathbf{H}}_{\mathrm{DL}}$, and
$\tilde{\mathbf{H}}_{\mathrm{DR}}$
used in the BSANN loss functions.

\subsection{Active XTC with Teacher-Protected PSZ Objective}
\label{sec:xtc}

The ear-wise BSANN described above is first trained purely for personal
sound zones using the loss in~\eqref{eq:lossPSZ}.  
For convenience, we define
\begin{equation}
\begin{aligned}
    &\mathcal{L}_{\mathrm{BZ}}      \equiv \mathcal{L}_1, 
    &&\mathcal{L}_{\mathrm{DZ}}      \equiv \mathcal{L}_2, \\
    &\mathcal{L}_{\mathrm{gain}}    \equiv \mathcal{L}_3, 
    &&\mathcal{L}_{\mathrm{compact}} \equiv \mathcal{L}_4.
\end{aligned}
\end{equation}

The \emph{PSZ pretraining} stage (Step~1 in Fig.~\ref{fig:BSANN}) minimizes
\begin{equation}
\label{eq:L_PSZ}
    \mathcal{L}_{\mathrm{PSZ}}
    = \alpha \mathcal{L}_{\mathrm{BZ}}
      + (1-\alpha)\mathcal{L}_{\mathrm{DZ}}
      + \beta \mathcal{L}_{\mathrm{gain}}
      + \gamma \mathcal{L}_{\mathrm{compact}},
\end{equation}
yielding a baseline BSANN that optimizes bright-zone and dark-zone performance
without explicit consideration of active XTC.

To enable loudspeaker-based binaural rendering with XTC, we introduce an XTC objective that penalizes interaural leakage while preserving the intended ear responses. For each frequency bin $\omega_n$, let
$\mathbf{P}(\omega_n)\in\mathbb{C}^{M\times 2\times L}$
denote the acoustic plant for the two bright-zone ears only, with $M$ control points per ear and $L$ loudspeakers.
Let
$\mathbf{W}(\omega_n)\in\mathbb{C}^{L\times 2}$
denote the BSANN-predicted stereo filters applied to these two target ears.
The resulting effective ear transfer matrix is
\begin{equation}
\label{eq:T_eff}
    \mathbf{T}_{\mathrm{eff}}(\omega_n)
    =
    \mathbf{P}(\omega_n)\,\mathbf{W}(\omega_n)
    \in \mathbb{C}^{M\times 2\times 2},
\end{equation}
whose diagonal entries $(R_{LL},R_{RR})$ correspond to the intended
bright-zone ear responses and whose off-diagonal entries $(R_{LR},R_{RL})$
represent interaural crosstalk.

\noindent\textit{1) Off-diagonal crosstalk loss:}
For each frequency bin $\omega_n$, the relative interaural leakage is defined as
\begin{equation}
\label{eq:xtc_r}
    r(\omega_n)
    =
    \frac{1}{2}
    \left[
        \frac{|R_{RL}(\omega_n)|^2}{|R_{LL}(\omega_n)|^2}
        +
        \frac{|R_{LR}(\omega_n)|^2}{|R_{RR}(\omega_n)|^2}
    \right].
\end{equation}
To prevent low-energy frequency regions from dominating the loss, we apply log-compression and an energy-based weighting. Define the diagonal energy
\begin{equation}
\label{eq:E_diag}
    E(\omega_n)
    =
    \frac{1}{2}\Big(
        |R_{LL}(\omega_n)|^2
        +
        |R_{RR}(\omega_n)|^2
    \Big),
\end{equation}
which is treated as a stop-gradient weight in the implementation. The off-diagonal XTC loss is computed as a normalized weighted average:
\begin{equation}
\label{eq:L_off}
    \mathcal{L}_{\mathrm{off}}
    =
    \frac{1}{N M}
    \sum_{n=1}^{N}
    \frac{
        E(\omega_n)\,
        \log\!\big(1+r(\omega_n)\big)
    }{
        \sum_{n'=1}^{N}E(\omega_{n'})
    }.
\end{equation}
This term penalizes crosstalk while suppressing the influence of frequency bins with very low intended ear energy.

\noindent\textit{2) Diagonal matching loss:}
To prevent the intended ear responses from degrading during the active XTC stage,
we match their magnitudes to a target diagonal response $\mathbf{T}_{\mathrm{target}}(\omega_n)$.  
For each frequency bin $\omega_n$, let
$R_{LL}^\mathrm{target}(\omega_n)$ and $R_{RR}^\mathrm{target}(\omega_n)$ denote the corresponding target magnitudes.  
The diagonal matching loss is defined as
\begin{equation}
\label{eq:L_diag}
\begin{aligned}
    \mathcal{L}_{\mathrm{diag}}
    &=
    \frac{1}{N M}
    \sum_{n=1}^{N}
    \frac{1}{2}
    \left[
        \left(
            \frac{|R_{LL}(\omega_n)|}{|R_{LL}^{\mathrm{target}}(\omega_n)|}
            - 1
        \right)^{2}
    \right. \\
    &\qquad\qquad\quad\left.
        +
        \left(
            \frac{|R_{RR}(\omega_n)|}{|R_{RR}^{\mathrm{target}}(\omega_n)|}
            - 1
        \right)^{2}
    \right].
\end{aligned}
\end{equation}
This term preserves the desired diagonal ear responses by penalizing deviations from the target magnitudes on a per-frequency basis.

\noindent\textit{3) Regularization via plant conditioning:}
To prevent unstable filter solutions during active XTC, we include an adaptive effort regularizer whose weight depends on the conditioning of the local plant Gram matrix~\cite{golub2013matrix}. For each frequency bin $\omega_n$ and each control point $m\in\{1,\dots,M\}$,
let $\mathbf{P}_m(\omega_n)\in\mathbb{C}^{2\times L}$ denote the ear-wise plant at that control point. The corresponding Gram matrix is
\begin{equation}
    \mathbf{G}_m(\omega_n)
    =
    \mathbf{P}_m(\omega_n)^{\mathrm{H}}\mathbf{P}_m(\omega_n)
    \in\mathbb{C}^{L\times L}.
\end{equation}
Let $\lambda_{\max,m}(\omega_n)$ and $\lambda_{\min,m}(\omega_n)$ denote the largest and smallest eigenvalues of $\mathbf{G}_m(\omega_n)$, and define the conditioning
ratio
\begin{equation}
    \kappa_m(\omega_n)
    =
    \frac{\lambda_{\max,m}(\omega_n)}{\lambda_{\min,m}(\omega_n)}.
\end{equation}
Following the implementation, the conditioning-dependent weight is
\begin{equation}
    \beta_m(\omega_n)
    =
    \beta_0\,
    \mathrm{ReLU}\!\left(
        \frac{\kappa_m(\omega_n)-\kappa_{\min}}{\kappa_{\min}}
    \right)
    \frac{\mathrm{tr}\big(\mathbf{G}_m(\omega_n)\big)}{L},
\end{equation}
where $\beta_0$ controls the base strength and $\kappa_{\min}$ determines
the threshold above which conditioning penalties are activated. The regularization loss is then computed as
\begin{equation}
\label{eq:L_reg}
    \mathcal{L}_{\mathrm{reg}}
    =
    \frac{1}{N M}
    \sum_{n=1}^{N}
    \sum_{m=1}^{M}
    \beta_m(\omega_n)\,
    \big\|\mathbf{W}(\omega_n)\big\|_F^{2},
\end{equation}
which penalizes large filter magnitudes more strongly at frequency--control point pairs where the plant is poorly conditioned.

\noindent\textit{4) Final XTC objective:}
The complete XTC objective combines the off-diagonal crosstalk penalty, the diagonal response matching term, and the conditioning-dependent effort regularizer:
\begin{equation}
\label{eq:L_XTC_final}
    \mathcal{L}_{\mathrm{XTC}}
    =
    \lambda_{\mathrm{off}}\,\mathcal{L}_{\mathrm{off}}
    +
    \lambda_{\mathrm{diag}}\,\mathcal{L}_{\mathrm{diag}}
    +
    \lambda_{\mathrm{reg}}\,\mathcal{L}_{\mathrm{reg}}.
\end{equation}

\noindent\textit{5) Teacher-protected adaptation:}
Directly optimizing~\eqref{eq:L_XTC_final} can substantially disturb the
pretrained PSZ solution and degrade the bright-zone and dark-zone energy
balance. To prevent such degradation of isolation performance, we introduce
a second training stage (Step~2 in Fig.~\ref{fig:BSANN}), in which the BSANN
is adapted from the PSZ-pretrained filters using a compound objective:
\begin{equation}
\label{eq:L_total}
\begin{aligned}
    \mathcal{L}_{\mathrm{total}}
    &= \lambda_{\mathrm{xtc}}\,\mathcal{L}_{\mathrm{XTC}}
       + w_{\mathrm{BZ}}\,\mathcal{L}_{\mathrm{BZ}}
       + w_{\mathrm{DZ}}\,\mathcal{L}_{\mathrm{DZ}} \\
    &\quad+ \beta\,\mathcal{L}_{\mathrm{gain}}
       + \gamma\,\mathcal{L}_{\mathrm{compact}}
       + \eta\,\mathcal{L}_{\mathrm{teach}}.
\end{aligned}
\end{equation}
Here, $\lambda_{\mathrm{xtc}}$ controls the relative importance of the XTC
objective, while $w_{\mathrm{BZ}}$ and $w_{\mathrm{DZ}}$ are small
\emph{protector} weights that preserve the original bright-zone and dark-zone behavior during active XTC.

The final term in~\eqref{eq:L_total} is a teacher-anchoring loss that keeps the updated filters close to the pretrained PSZ solution.  
Let $\mathbf{F}_{\mathrm{cur}}(\omega_n)$ and $\mathbf{F}_{\mathrm{teach}}(\omega_n)$
denote the stacked frequency-domain filters (all loudspeakers and program channels) of the current and pretrained BSANN at frequency $\omega_n$:
\begin{equation}
\begin{aligned}
\mathbf{F}_{\mathrm{cur}}(\omega_n)
    &= \mathrm{vec}\big(\mathbf{W}_{\mathrm{cur}}(\omega_n)\big),\quad
    \\
\mathbf{F}_{\mathrm{teach}}(\omega_n)
    &= \mathrm{vec}\big(\mathbf{W}_{\mathrm{teach}}(\omega_n)\big).
\end{aligned}
\end{equation}
The teacher-anchoring loss is then defined as
\begin{equation}
\label{eq:L_teacher}
    \mathcal{L}_{\mathrm{teach}}
    = \frac{1}{N}
      \sum_{n=1}^{N}
      \big\|
        \mathbf{F}_{\mathrm{cur}}(\omega_n)
        - \mathbf{F}_{\mathrm{teach}}(\omega_n)
      \big\|_2^{2},
\end{equation}
which discourages large deviations from the pretrained PSZ filters while still allowing the network to adapt in directions that improve
$\mathcal{L}_{\mathrm{XTC}}$.

In practice, \(w_{\mathrm{BZ}}\) and \(w_{\mathrm{DZ}}\) are set to values smaller than 
\(\lambda_{\mathrm{xtc}}\), so that the bright-zone and dark-zone terms act only as 
weak regularizers that preserve the original PSZ field during the active XTC stage.
The teacher-anchoring term $\mathcal{L}_{\mathrm{teach}}$ provides an additional global constraint on filter shape, enabling stable training
during the XTC stage. The resulting two-stage procedure, consisting of PSZ pretraining with
$\mathcal{L}_{\mathrm{PSZ}}$ followed by active XTC with
$\mathcal{L}_{\mathrm{total}}$, produces filters that simultaneously achieve
high IZI, IPI, and strong XTC, as demonstrated in Sec.~\ref{sec:results}.
\section{Experimental Setup and Evaluation}
\label{sec:experiment}

This section describes the experimental setup used to evaluate the proposed BSANN framework, including the loudspeaker configuration, listener geometry,
training dataset, model parameters, and evaluation metrics. Experiments were designed to quantify IZI, IPI, and XTC under realistic acoustic conditions.

\subsection{Loudspeaker Layout and Listener Geometry}

\begin{figure}[!t]
    \centering
    \includegraphics[width=\linewidth]{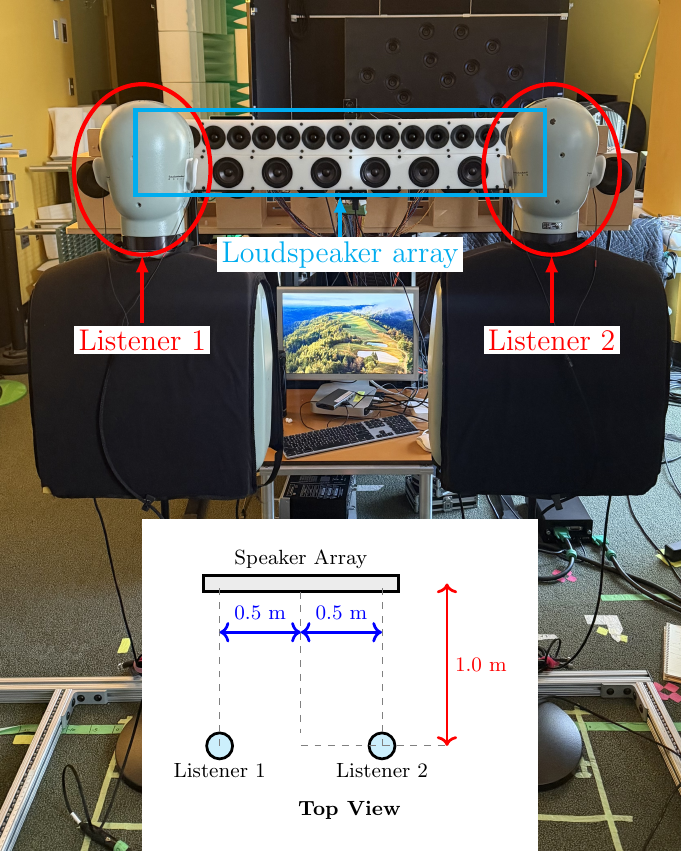}
    \caption{Physical experimental setup. 
The loudspeaker system consists of a 24-element array mounted at listener height, 
with 8 woofers in the lower row and 16 tweeters in the upper row. 
Two head-and-torso simulators (B\&K HATS) were positioned symmetrically about 
the array center, each placed 0.5~m to the left or right and 1.0~m in front of 
the loudspeaker array, as illustrated in the top-view diagram.}
    \label{fig:system}
\end{figure}

Figure~\ref{fig:system} shows the physical setup used for evaluation.
The 24-loudspeaker array is mounted on a rigid baffle at listener height
and arranged in two horizontal rows: 8 woofers (100--2000~Hz) in the lower row
and 16 tweeters (2--20~kHz) in the upper row. The drivers are spaced uniformly along a horizontal arc and face toward the listeners. The array geometry and driver spacing match the configuration assumed during BSANN training.

Two head-and-torso simulators (B\&K HATS) were positioned symmetrically about
the array centerline, each located 0.5~m to the left or right of the center
and 1.0~m in front of the array. Both HATS were aligned at the same height
as the loudspeakers, with their interaural axis perpendicular to the array
and centered relative to its midpoint.

\subsection{Training Dataset and Procedure}

The training dataset was generated using simulated room impulse responses
(RIRs) based on the shoebox-room procedure of the original SANN-PSZ
work~\cite{qiao2025sann}. For each training instance, random room
dimensions, reverberation time, array position, and listener zone centers
were drawn, and gpuRIR~\cite{DiazGuerra2021gpuRIR} was used to simulate RIRs from all loudspeakers to
the control points. In addition to the randomization strategy used in
\cite{qiao2025sann}, the present dataset incorporates variability
in the listener geometry: the rigid-sphere head radius was sampled
uniformly within a realistic range, the ear positions were offset by a
small random distance from the sphere surface to account for pinna depth,
and the ear-adjacent control points were obtained by uniformly sampling a
circular region around each ear while discarding points lying inside the
rigid-sphere boundary.

Using the simulated RIRs, physically informed acoustic transfer functions
(ATFs) were then constructed following Sec.~\ref{sec:physmodel}. Each
ATF $\tilde{H}_{e,m,\ell}(\omega)$ combines the simulated direct and
reflected room responses with the measured anechoic loudspeaker response
$A_\ell(\omega)$, analytic piston directivity $D_\ell(\omega,\theta_{e,m,\ell})$,
and the rigid-sphere HRTF $H_{\mathrm{HRTF},e,m}(\omega)$. These
physically informed ATFs form the full training dataset used in the PSZ
pretraining and subsequent active XTC stages.

Using these physically informed ATFs, the BSANN was trained in two stages.
In the first stage, the network was pretrained for personal sound zones using the loss function defined in Eq.~\eqref{eq:L_PSZ}, with hyperparameters
$\alpha=0.5$, $\beta=0.5$, and $\gamma=0.5$.

In the second stage, the model was refined according to the composite objective 
dominated by the XTC loss, as defined in Eq.~\eqref{eq:L_total}. The 
hyperparameters were selected based on an extensive parametric search aimed at 
maximizing both isolation and XTC performance. Because the resulting 
performance was found to be only weakly sensitive to the precise hyperparameter 
values, we did not systematize the search or report it in detail. The adopted 
values for this study are $\lambda_{\mathrm{xtc}}=0.14$, $\lambda_{\mathrm{off}}=1.5$, 
$\lambda_{\mathrm{diag}}=1.0$, $\lambda_{\mathrm{reg}}=1.0$, 
$\beta_{0}=10^{-4}$, $\kappa_{\min}=10^{3}$, and $\varepsilon=10^{-8}$. 
The bright-zone and dark-zone protector weights were set to 
$w_{\mathrm{BZ}}=w_{\mathrm{DZ}}=0.075$. Both the gain and compactness weights 
were kept at their original values, $\beta=0.5$ and $\gamma=0.5$, respectively.
The teacher-anchoring coefficient was set to $\eta=1.0$.

\subsection{Evaluation Metrics}
\label{sec:metrics}
We adopt three objective metrics for evaluation: inter-zone isolation
(IZI), inter-program isolation (IPI), and crosstalk cancellation (XTC)~\cite{qiao2022isolation,qiao2025sann,choueiri2018binaural}. 
Let
$\tilde{\mathbf{H}}_{1}(\omega)
=[\mathbf{H}_{1,\mathrm{L}}(\omega),\, \mathbf{H}_{1,\mathrm{R}}(\omega)]^{\mathsf{T}}$
denote the ear-wise ATF submatrix for Listener~1, and similarly let
$\tilde{\mathbf{H}}_{2}(\omega)
=[\mathbf{H}_{2,\mathrm{L}}(\omega),\, \mathbf{H}_{2,\mathrm{R}}(\omega)]^{\mathsf{T}}$ 
denote the corresponding submatrix for Listener~2.
Let 
$\tilde{\mathbf{g}}_{1}(\omega)
=[\mathbf{g}_{1,\mathrm{L}}(\omega),\, \mathbf{g}_{1,\mathrm{R}}(\omega)]^{\mathsf{T}}$
denote the BSANN-generated stereo filters for Program~1, and
$\tilde{\mathbf{g}}_{2}(\omega)
=[\mathbf{g}_{2,\mathrm{L}}(\omega),\, \mathbf{g}_{2,\mathrm{R}}(\omega)]^{\mathsf{T}}$
those for Program~2.

IZI evaluates the acoustic separation between the
bright and dark zones associated with two different listeners. The IZI is
defined as~\cite{qiao2025sann,qiao2022isolation}
\begin{equation}
\begin{aligned}
\mathrm{IZI}_1(\omega)
&=
10\log_{10}
\frac{
\|\tilde{\mathbf{H}}_{1}(\omega)\tilde{\mathbf{g}}_{1}(\omega)\|_2^2
}{
\|\tilde{\mathbf{H}}_{2}(\omega)\tilde{\mathbf{g}}_{1}(\omega)\|_2^2
},
\\
\mathrm{IZI}_2(\omega)
&=
10\log_{10}
\frac{
\|\tilde{\mathbf{H}}_{2}(\omega)\tilde{\mathbf{g}}_{2}(\omega)\|_2^2
}{
\|\tilde{\mathbf{H}}_{1}(\omega)\tilde{\mathbf{g}}_{2}(\omega)\|_2^2
}.
\end{aligned}
\label{eq:IZI_metric}
\end{equation}

Similarly, IPI evaluates the interference between two independent program channels rendered to the same listener. The IPI is defined as~\cite{qiao2025sann,qiao2022isolation}
\begin{equation}
\begin{aligned}
\mathrm{IPI}_1(\omega)
&=
10\log_{10}
\frac{
\|\tilde{\mathbf{H}}_{1}(\omega)\tilde{\mathbf{g}}_{1}(\omega)\|_2^2
}{
\|\tilde{\mathbf{H}}_{1}(\omega)\tilde{\mathbf{g}}_{2}(\omega)\|_2^2
},
\\
\mathrm{IPI}_2(\omega)
&=
10\log_{10}
\frac{
\|\tilde{\mathbf{H}}_{2}(\omega)\tilde{\mathbf{g}}_{2}(\omega)\|_2^2
}{
\|\tilde{\mathbf{H}}_{2}(\omega)\tilde{\mathbf{g}}_{1}(\omega)\|_2^2
}.
\end{aligned}
\label{eq:IPI_metric}
\end{equation}

XTC evaluates the suppression of contralateral
leakage between the two ears of the active listener. For the two possible
program assignments, let
$\mathbf{T}_{\mathrm{eff},1}(\omega_n)$ and
$\mathbf{T}_{\mathrm{eff},2}(\omega_n)$
denote the effective ear-transfer matrices for Program~1 and Program~2,
respectively. The diagonal entries $(R_{LL,k},R_{RR,k}), (k\in\{1,2\})$ represent the intended
ipsilateral responses and the off-diagonal entries
$(R_{LR,k},R_{RL,k})$ represent residual crosstalk. The two XTC measures are defined as~\cite{choueiri2018binaural}
\begin{equation}
\begin{aligned}
\mathrm{XTC}_1(\omega)
&=
10\log_{10}
\frac{|R_{LL,1}(\omega)|^2 + |R_{RR,1}(\omega)|^2}
{|R_{LR,1}(\omega)|^2 + |R_{RL,1}(\omega)|^2},
\\[4pt]
\mathrm{XTC}_2(\omega)
&=
10\log_{10}
\frac{|R_{LL,2}(\omega)|^2 + |R_{RR,2}(\omega)|^2}
{|R_{LR,2}(\omega)|^2 + |R_{RL,2}(\omega)|^2}.
\end{aligned}
\label{eq:XTC_metric}
\end{equation}

\subsection{Test Conditions}

Experiments were designed to evaluate the impact of stereo modeling,
acoustic realism in the ATF construction, and active XTC on overall
PSZ and binaural performance. Three sets of conditions were considered:

\begin{itemize}
\item \textbf{SANN vs.\ BSANN:}
Comparison between the monophonic SANN-PSZ model and a BSANN-PSZ model trained 
with ideal point-source ATFs and without the physically informed components or the 
active XTC stage, so that the gains attributable solely to ear-wise control can be quantified.

\item \textbf{Simulated point-source ATFs vs.\ physically informed ATFs:}
Comparison between a BSANN-PSZ model trained with ideal point-source ATFs and a 
BSANN-PSZ model trained with physically informed ATFs that incorporate the measured 
loudspeaker responses, analytic piston directivity, and rigid-sphere HRTFs, with 
both models trained without the active XTC stage, so that the benefits attributable 
to physically informed acoustic modeling can be quantified.

\item \textbf{With vs.\ without active XTC:}
Comparison between a BSANN-PSZ model trained without the active XTC stage and a 
BSANN-PSZ model trained with the active XTC stage. Both models were trained using the 
same physically informed ATFs, so that the contribution of the active XTC 
stage to overall performance can be quantified.

\end{itemize}

All conditions were evaluated under static head poses. For evaluation, broadband 
swept-sine signals were passed through each model’s filters and reproduced over 
the loudspeaker array. The resulting ear signals for Listener~1 and Listener~2 
(both represented by B\&K HATS) were recorded using miniature in-ear microphones 
(Theoretica Applied Physics' BACCH-BM Pro Mk III) placed at the entrances of 
the left and right ear canals of each listener. These measured ear signals were 
then used to compute the IZI, IPI, and XTC metrics. The study was conducted in a 
listening room with asymmetric boundaries that is more representative of real 
listening spaces.

\section{Experimental Results}
\label{sec:results}

This section reports the measured performance of the proposed BSANN-PSZ model under the evaluation conditions described in Sec.~\ref{sec:experiment}. The
results are presented in terms of IZI, IPI, and XTC, using the metrics defined in Sec.~\ref{sec:metrics}. Unless otherwise noted,
all results are reported as log-frequency–weighted averages over 100--20{,}000~Hz.
\subsection{Comparison Between SANN-PSZ and BSANN-PSZ}

\begin{figure*}[!t]
\centering
\setlength{\tabcolsep}{2pt}

\begin{tabular}{ccc}
\subfloat[Listener~1 IZI]{\includegraphics[width=0.32\textwidth]{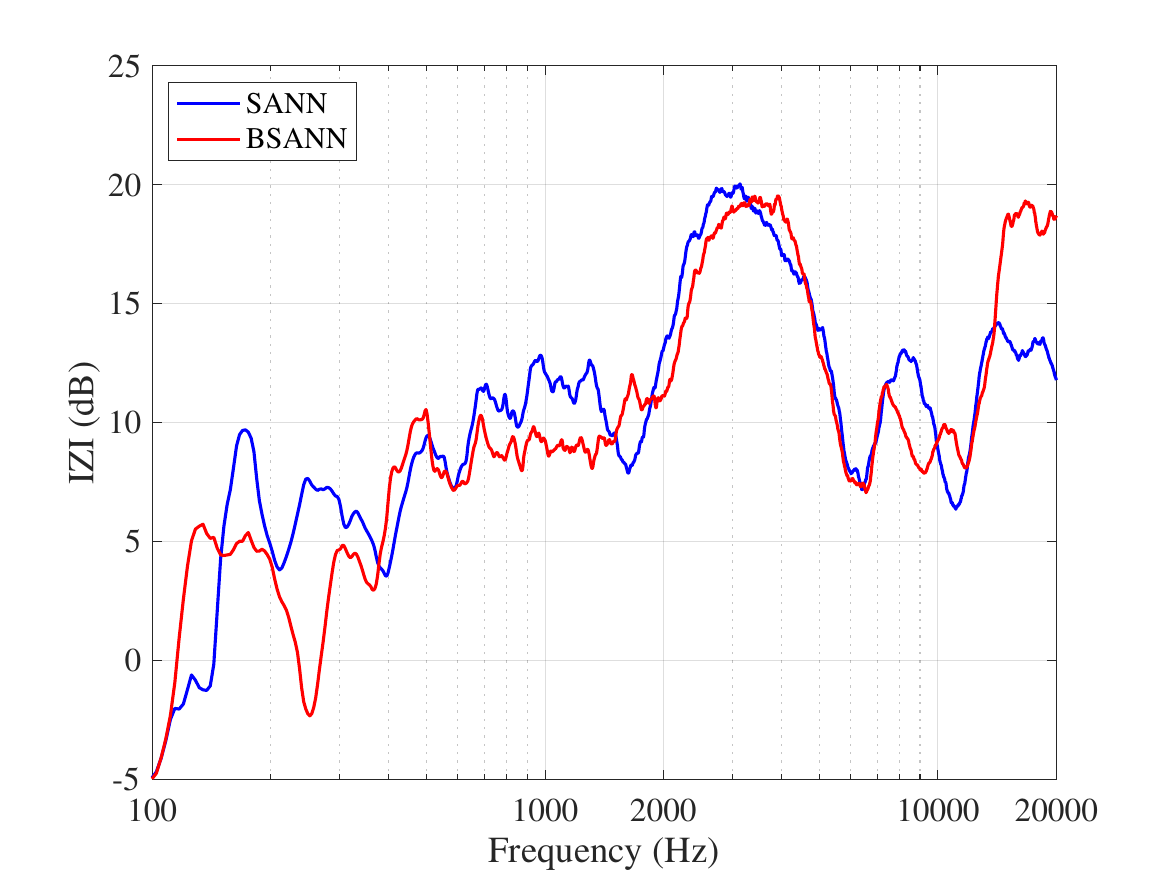}} &
\subfloat[Listener~1 IPI]{\includegraphics[width=0.32\textwidth]{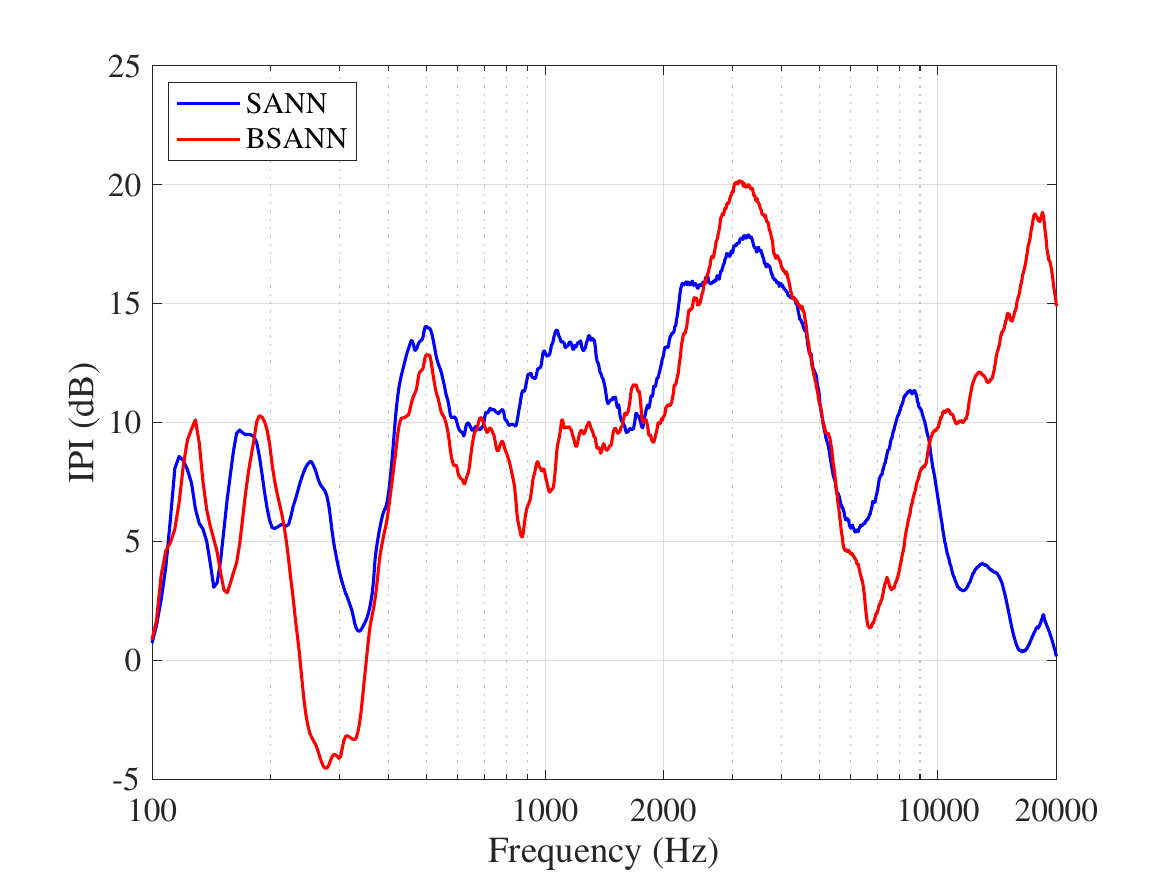}} &
\subfloat[Listener~1 XTC]{\includegraphics[width=0.32\textwidth]{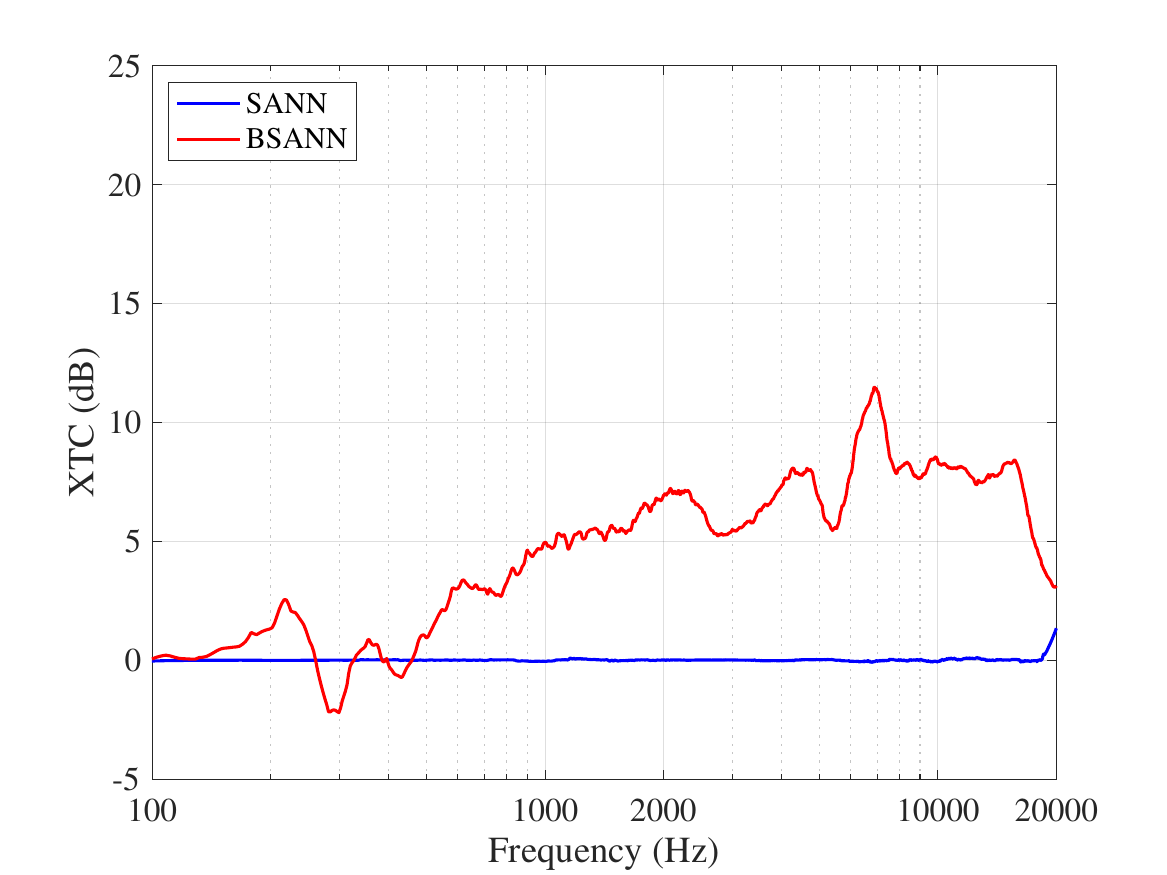}} \\[0pt]
\subfloat[Listener~2 IZI]{\includegraphics[width=0.32\textwidth]{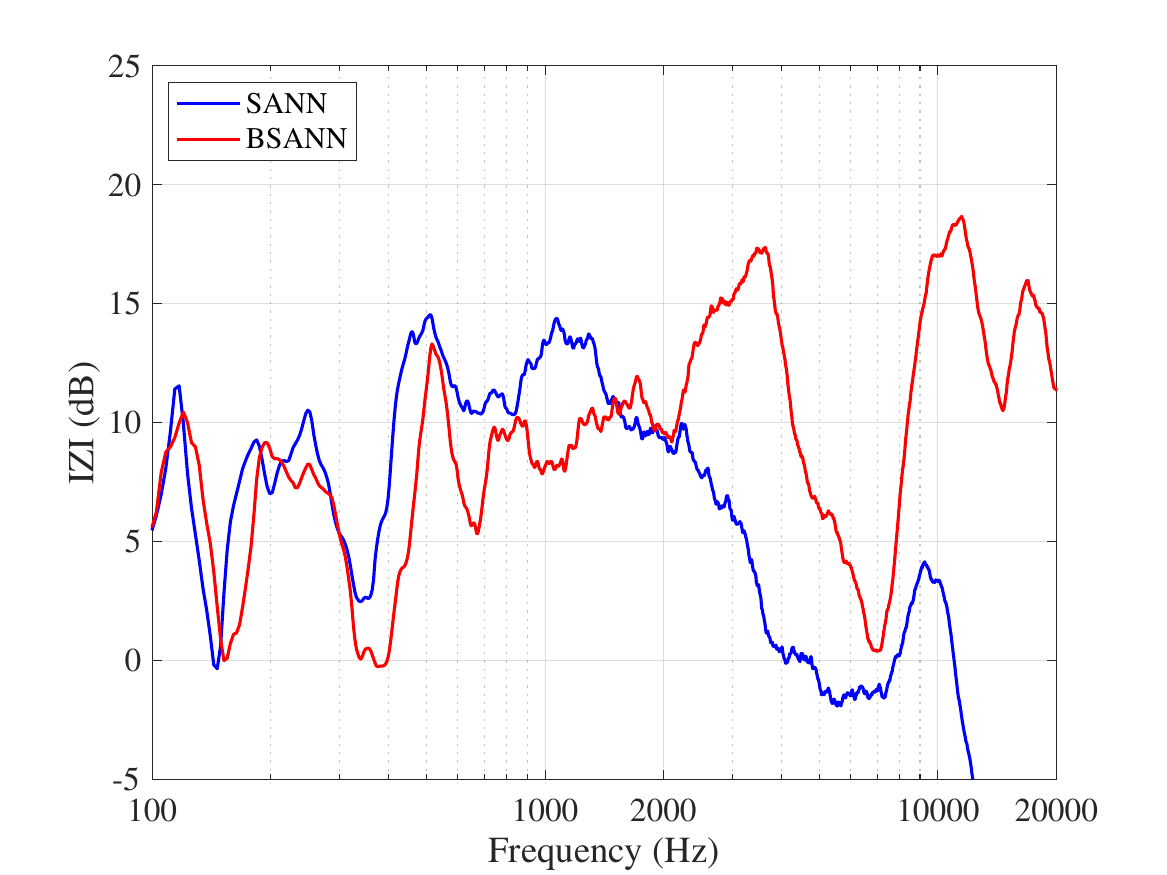}} &
\subfloat[Listener~2 IPI]{\includegraphics[width=0.32\textwidth]{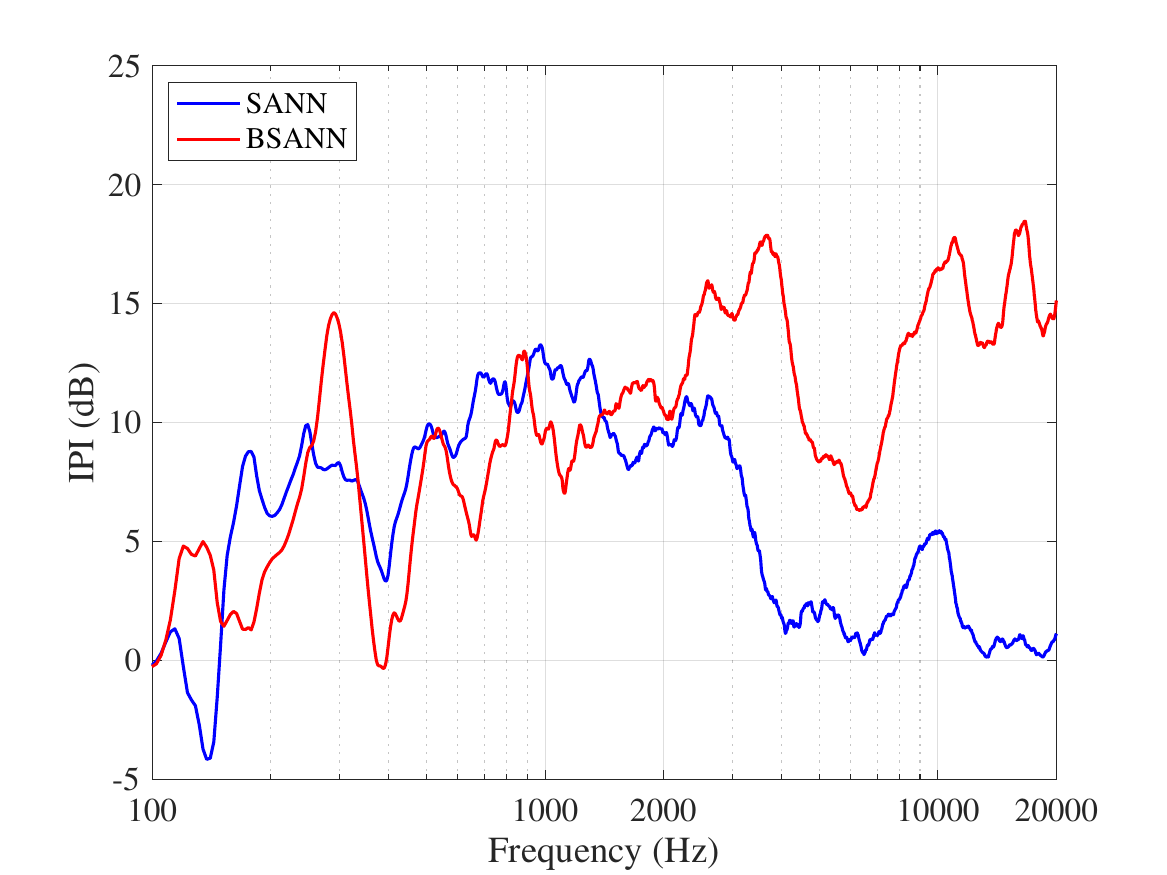}} &
\subfloat[Listener~2 XTC]{\includegraphics[width=0.32\textwidth]{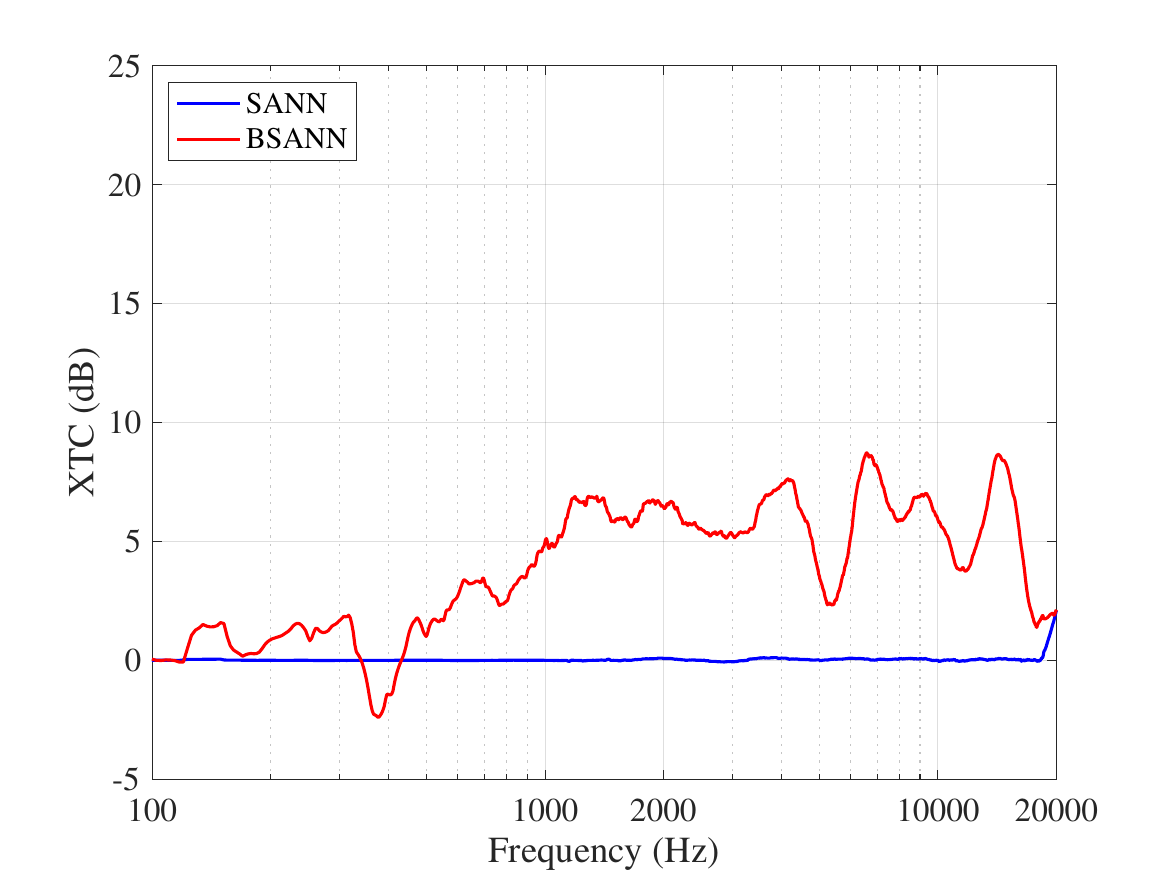}}
\end{tabular}

\caption{
Comparison between the monophonic SANN-PSZ model (blue) and the BSANN-PSZ model 
trained with ideal point-source ATFs and without physically informed components 
or active XTC (red). Each panel shows the log-frequency–weighted IZI, IPI, or XTC curve 
for Listener~1 and Listener~2.
}
\label{fig:sann_BSANN}
\end{figure*}

\begin{table}[!t]
\caption{Log-frequency–weighted mean IZI, IPI, and XTC for the monophonic SANN-PSZ model 
and the BSANN-PSZ model trained with ideal point-source ATFs and without the 
physically informed components or the active XTC stage. Subscripts 1 and 2 denote Listener~1 and Listener~2, respectively. The $\Delta$ column denotes the metric difference (BSANN minus SANN).\label{tab:sann_BSANN_means}}
\centering
\begin{tabular}{|c|c|c|c|}
\hline
Metric & SANN & BSANN & $\Delta$ \\
\hline
$\mathrm{IZI}_{1}$ [dB] & 9.99 & 9.54 & $-0.45$ \\
\hline
$\mathrm{IZI}_{2}$ [dB] & 5.11 & 9.18 & $+4.07$ \\
\hline
$\mathrm{IPI}_{1}$ [dB] & 9.18 & 8.98 & $-0.20$ \\
\hline
$\mathrm{IPI}_{2}$ [dB] & 5.92 & 9.74 & $+3.82$ \\
\hline
$\mathrm{XTC}_{1}$ [dB] & 0.00 & 4.97 & $+4.97$ \\
\hline
$\mathrm{XTC}_{2}$ [dB] & 0.01 & 4.38 & $+4.37$ \\
\hline
\end{tabular}
\end{table}

The log-frequency–weighted mean IZI, IPI, and XTC values for SANN and BSANN are 
summarized in Table~\ref{tab:sann_BSANN_means}, with the corresponding curves 
shown in Fig.~\ref{fig:sann_BSANN}. These results reveal that the monophonic 
formulation yields highly imbalanced performance across the two listeners: 
while Listener~1 achieves reasonable isolation, Listener~2 suffers a deficit of 
approximately 4–5~dB in both IZI and IPI. This imbalance is primarily caused by 
the asymmetric reflective environment on the right side of the room, where 
strong early reflections degrade the isolation at the Listener~2 position.

BSANN largely removes this imbalance. Relative to the SANN baseline, BSANN 
increases Listener~2 isolation by 4.07~dB in IZI and 3.82~dB in IPI, essentially 
eliminating the performance gap between the two listeners. For Listener~1, by 
contrast, the changes are minimal ($-$0.45~dB in IZI and $-$0.20~dB in IPI), 
demonstrating that the stereo ear-wise formulation preserves SANN’s performance 
where the monophonic model already works well while substantially improving 
performance at the disadvantaged listener.

The benefits of stereo ear-wise control are even more prominent in the 
XTC metric. For SANN, the log-frequency–weighted XTC is effectively zero for 
both programs, whereas BSANN increases XTC by 4.97~dB and 4.37~dB for 
Listener~1 and Listener~2, respectively. These passive XTC gains, obtained 
without the active XTC stage, highlight the inherent crosstalk-reducing 
capability of the stereo ear-wise formulation.

Together, these improvements show that the BSANN-PSZ architecture not only 
balances performance across listeners in an asymmetric room, but also provides 
substantial isolation and crosstalk gains that are unattainable with the 
monophonic SANN-PSZ architecture.

\subsection{Effect of Physically Informed ATFs}

\begin{figure*}[!t]
\centering
\setlength{\tabcolsep}{2pt}

\begin{tabular}{ccc}
\subfloat[Listener~1 IZI]{\includegraphics[width=0.32\textwidth]{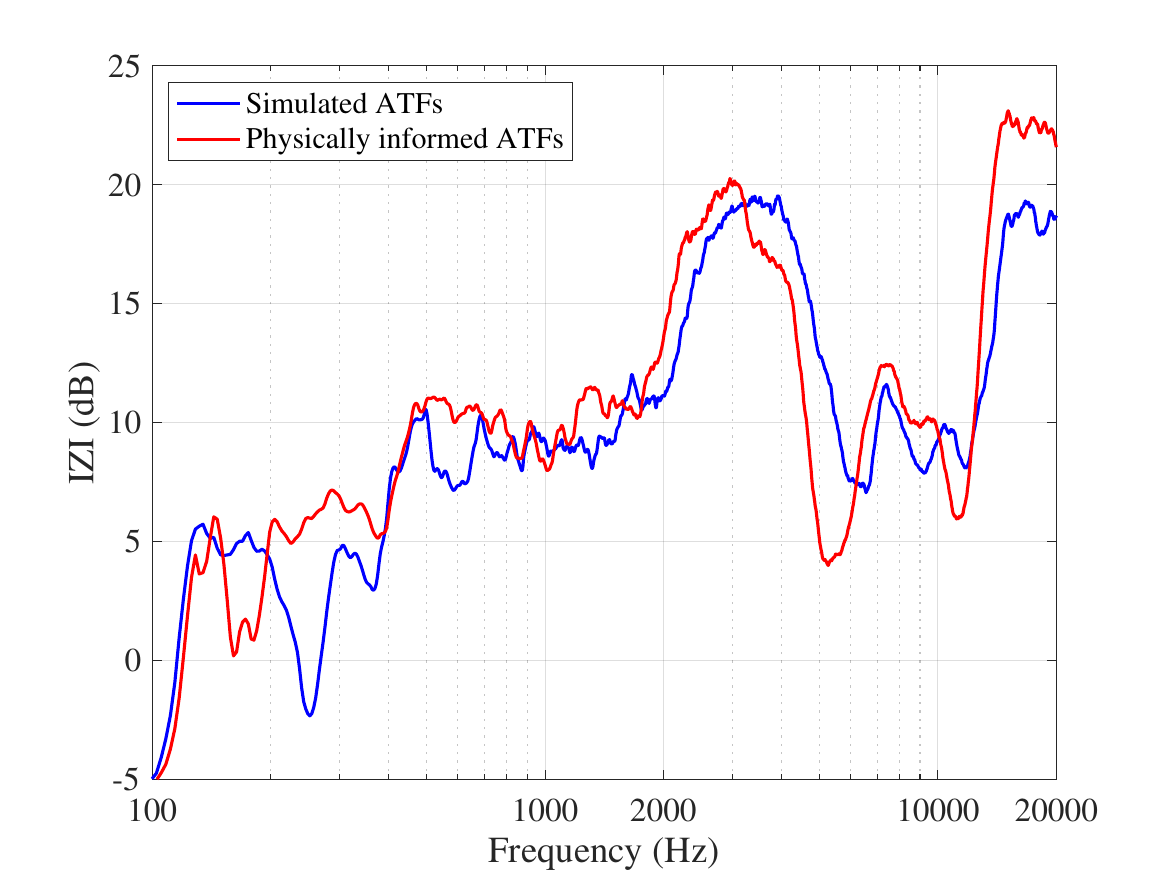}} &
\subfloat[Listener~1 IPI]{\includegraphics[width=0.32\textwidth]{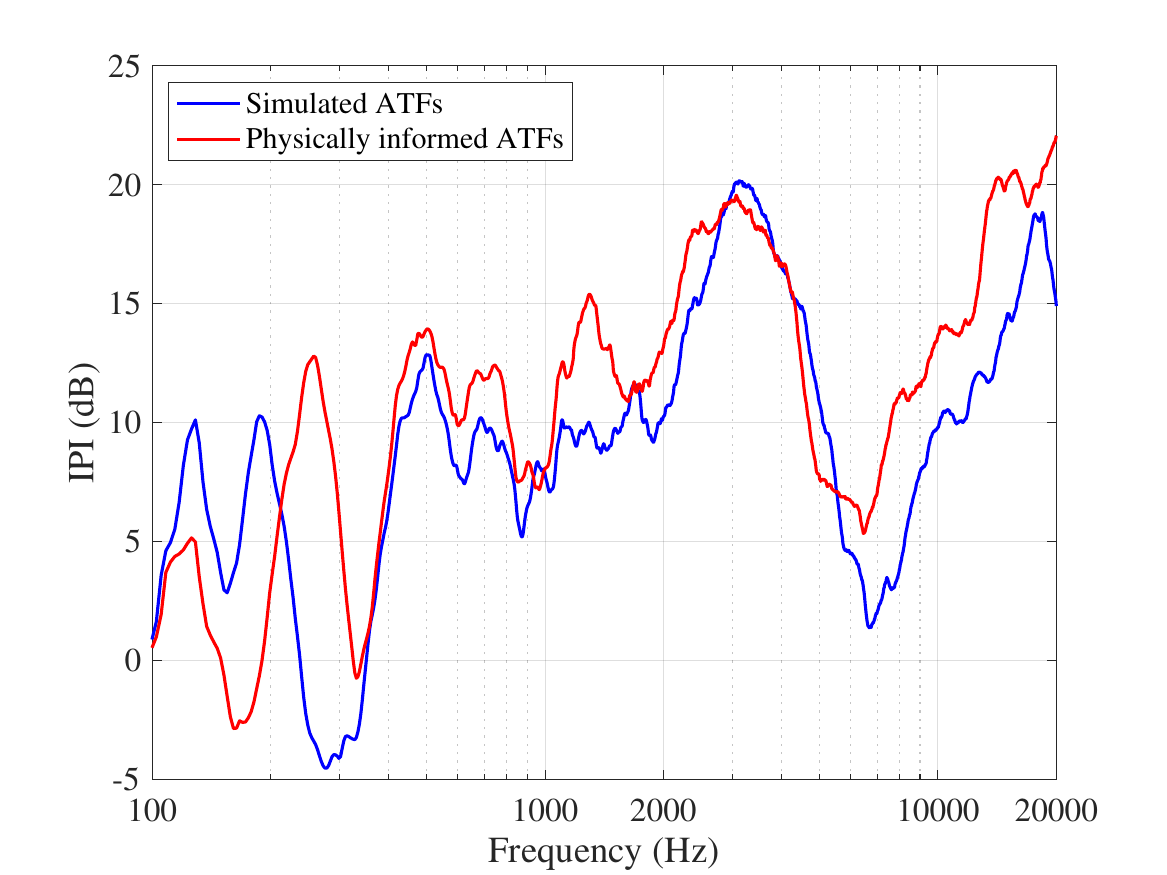}} &
\subfloat[Listener~1 XTC]{\includegraphics[width=0.32\textwidth]{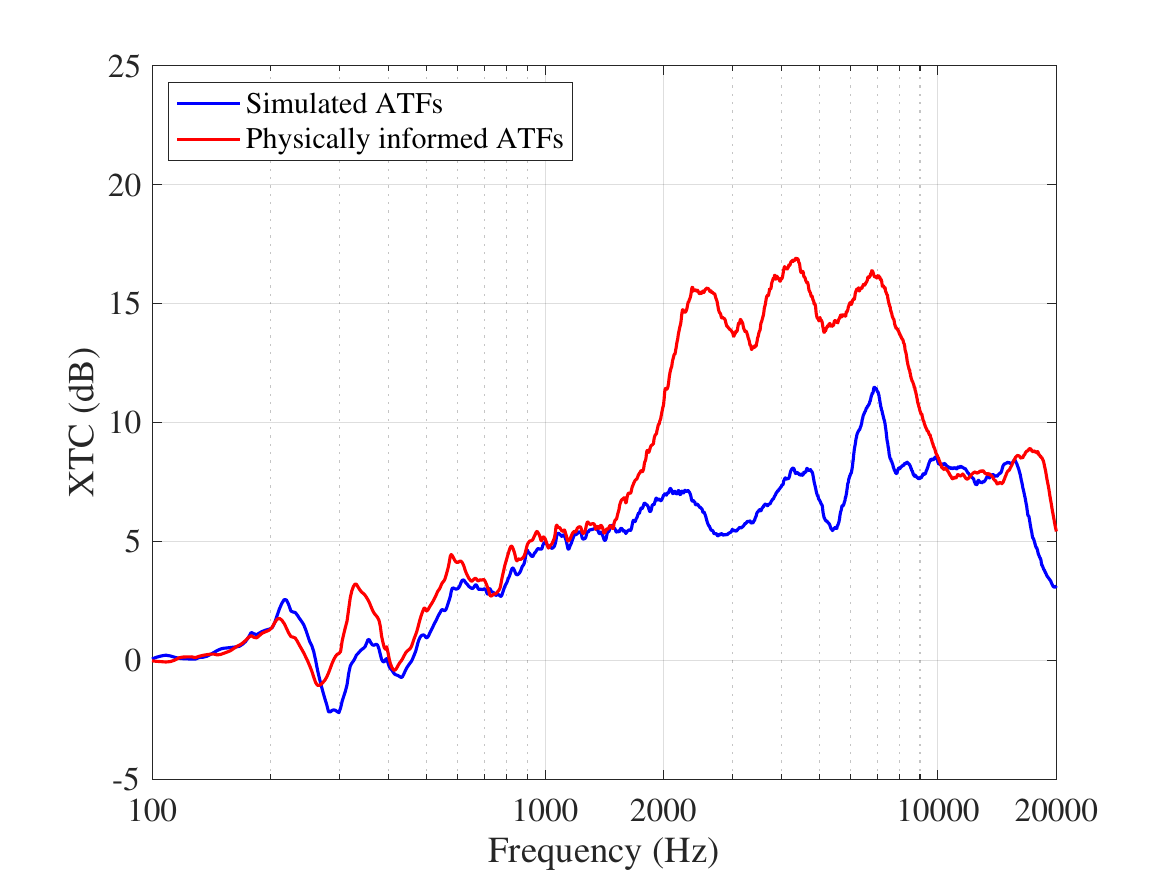}} \\[0pt]
\subfloat[Listener~2 IZI]{\includegraphics[width=0.32\textwidth]{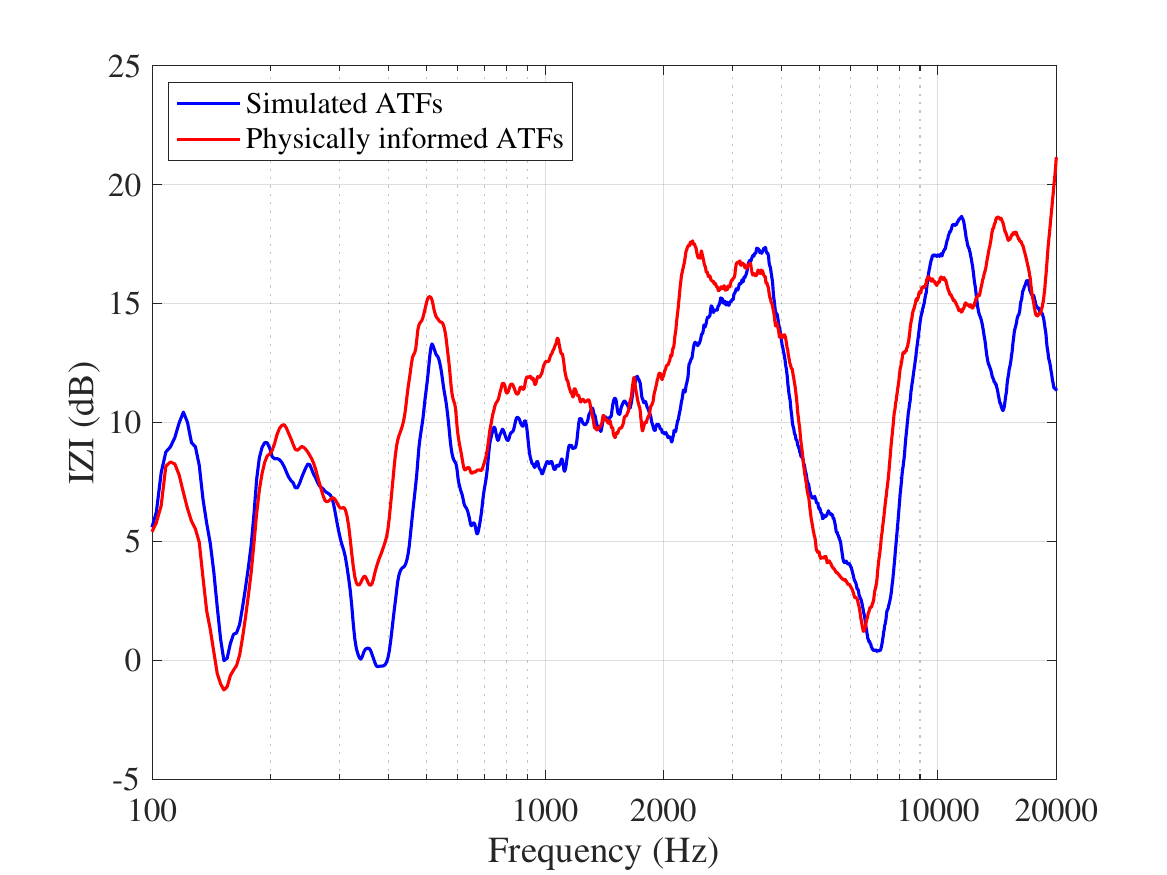}} &
\subfloat[Listener~2 IPI]{\includegraphics[width=0.32\textwidth]{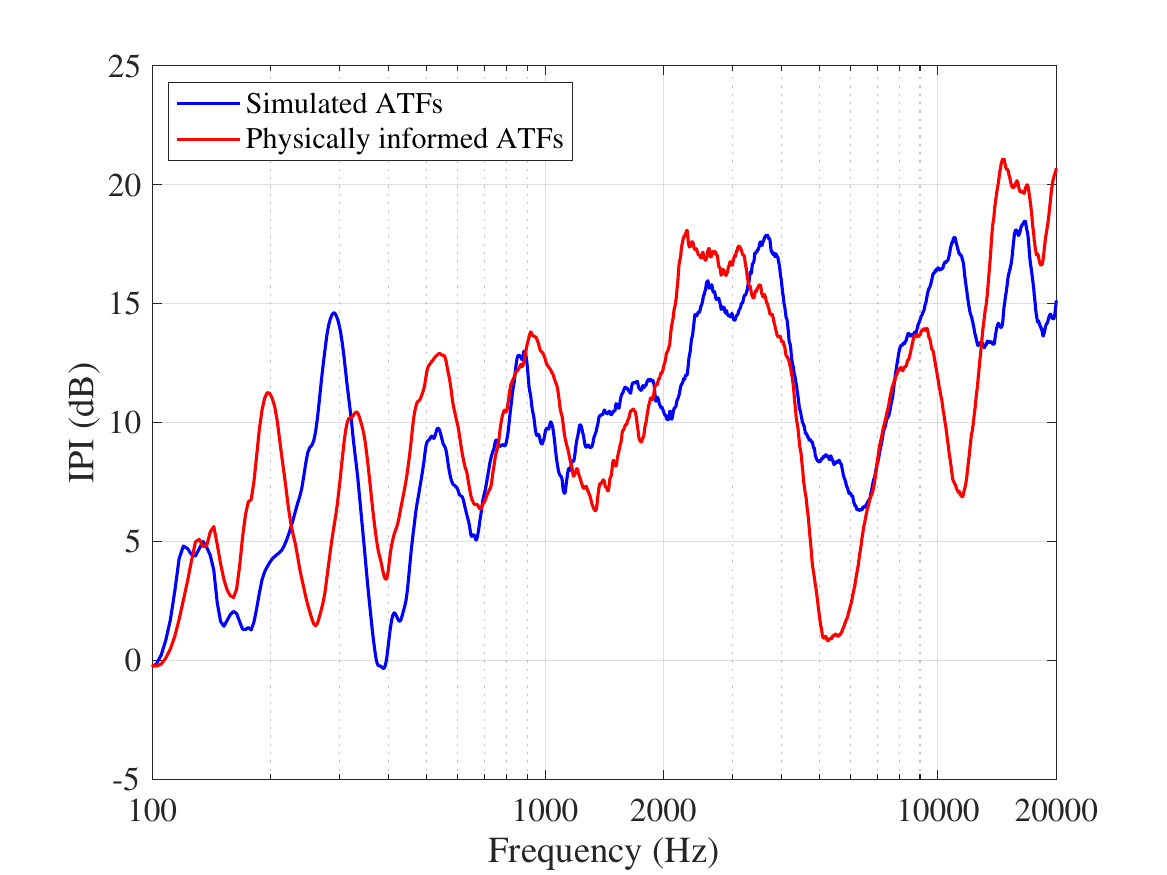}} &
\subfloat[Listener~2 XTC]{\includegraphics[width=0.32\textwidth]{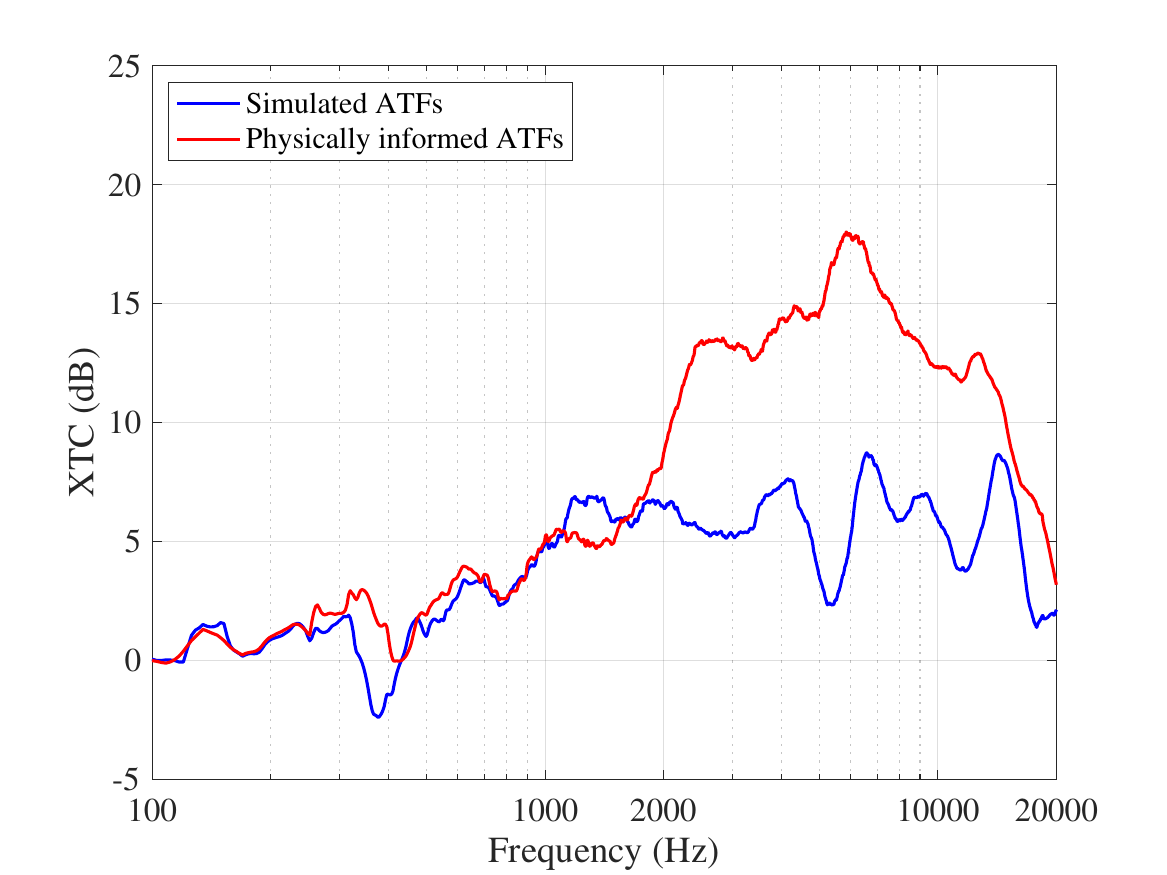}}
\end{tabular}

\caption{
Comparison between a BSANN-PSZ model trained with ideal point-source ATFs (blue) 
and a BSANN-PSZ model trained with physically informed ATFs (red), with both 
models trained without active XTC. Each panel shows the log-frequency–weighted 
IZI, IPI, or XTC curve for Listener~1 and Listener~2.
}
\label{fig:physATF_results}
\end{figure*}

\begin{table}[!t]
\caption{Log-frequency--weighted mean IZI, IPI, and XTC for BSANN-PSZ models 
trained with simulated point-source ATFs and with physically informed ATFs, both 
without active XTC. Subscripts 1 and 2 denote Listener~1 and Listener~2, 
respectively. The $\Delta$ column denotes the metric difference (Phys.\ informed 
minus Simulated).\label{tab:physATF_means}}
\centering
\begin{tabular}{|c|c|c|c|}
\hline
Metric & Simulated ATFs & Phys.\ informed ATFs & $\Delta$ \\
\hline
$\mathrm{IZI}_{1}$~[dB] & 9.54 & 10.23 & $+0.69$ \\
\hline
$\mathrm{IZI}_{2}$~[dB] & 9.18 & 10.43 & $+1.25$ \\
\hline
$\mathrm{IPI}_{1}$~[dB] & 8.98 & 10.87 & $+1.89$ \\
\hline
$\mathrm{IPI}_{2}$~[dB] & 9.74 & 9.79 & $+0.05$ \\
\hline
$\mathrm{XTC}_{1}$~[dB] & 4.97 & 7.93 & $+2.96$ \\
\hline
$\mathrm{XTC}_{2}$~[dB] & 4.38 & 8.19 & $+3.81$ \\
\hline
\end{tabular}
\end{table}

Figure~\ref{fig:physATF_results} and Table~\ref{tab:physATF_means} summarize the 
effect of physically informed acoustic modeling on BSANN performance. 
When trained with simulated point-source ATFs, BSANN already provides balanced isolation across the two listeners, achieving log-frequency–weighted values of 
$\mathrm{IZI}_1 = 9.54$~dB, $\mathrm{IZI}_2 = 9.18$~dB, 
$\mathrm{IPI}_1 = 8.98$~dB, and $\mathrm{IPI}_2 = 9.74$~dB.

Incorporating physically informed ATFs, consisting of measured loudspeaker 
responses, analytic piston directivity, and rigid-sphere HRTFs, yields 
isolation improvements for both listeners.  Relative to the simulated ATFs, the physically informed ATFs increase 
$\mathrm{IZI}_1$ and $\mathrm{IZI}_2$ by 0.69~dB and 1.25~dB, and increase 
$\mathrm{IPI}_1$ and $\mathrm{IPI}_2$ by 1.89~dB and 0.05~dB, respectively. These 
gains indicate that the added acoustic realism strengthens the ear-wise isolation 
achieved by BSANN.

The most pronounced differences appear in the XTC metric. With simulated ATFs, the 
log-frequency–weighted XTC levels are 4.97~dB and 4.38~dB for Listener~1 and 
Listener~2, respectively. Using physically informed ATFs increases these values to 
7.93~dB and 8.19~dB, corresponding to gains of 2.96~dB and 3.81~dB. These results 
indicate that including analytic piston directivity and ear-dependent 
scattering effects strengthens interaural separation and increases the fidelity of 
transmitting binaural (ITD, ILD) cues.

Overall, physically informed modeling enhances BSANN performance under 
realistic acoustic conditions by improving isolation performance and 
providing stronger passive XTC through more accurate loudspeaker frequency 
responses, radiation patterns, and ear-dependent scattering cues.

\subsection{Effect of Active XTC Stage}

\begin{figure*}[!t]
\centering
\setlength{\tabcolsep}{2pt}

\begin{tabular}{ccc}
\subfloat[Listener~1 IZI]{\includegraphics[width=0.32\textwidth]{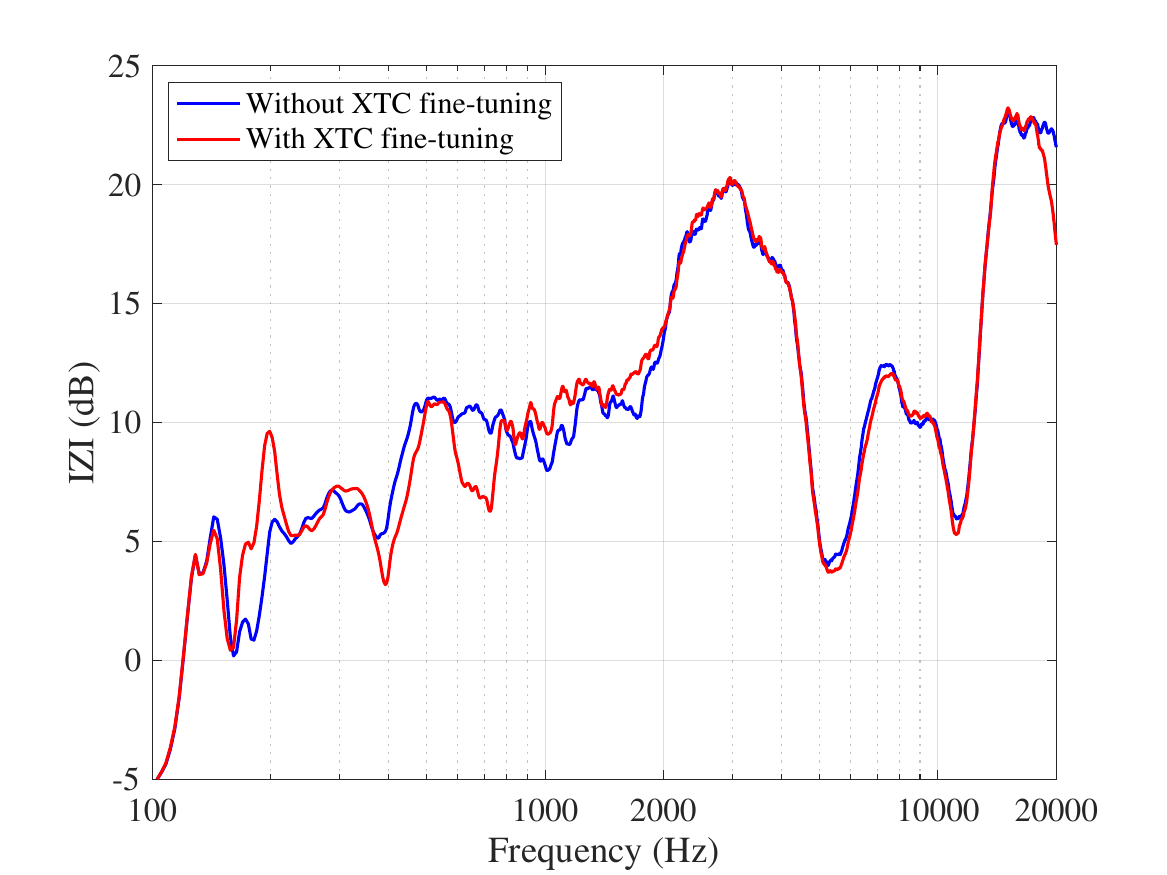}} &
\subfloat[Listener~1 IPI]{\includegraphics[width=0.32\textwidth]{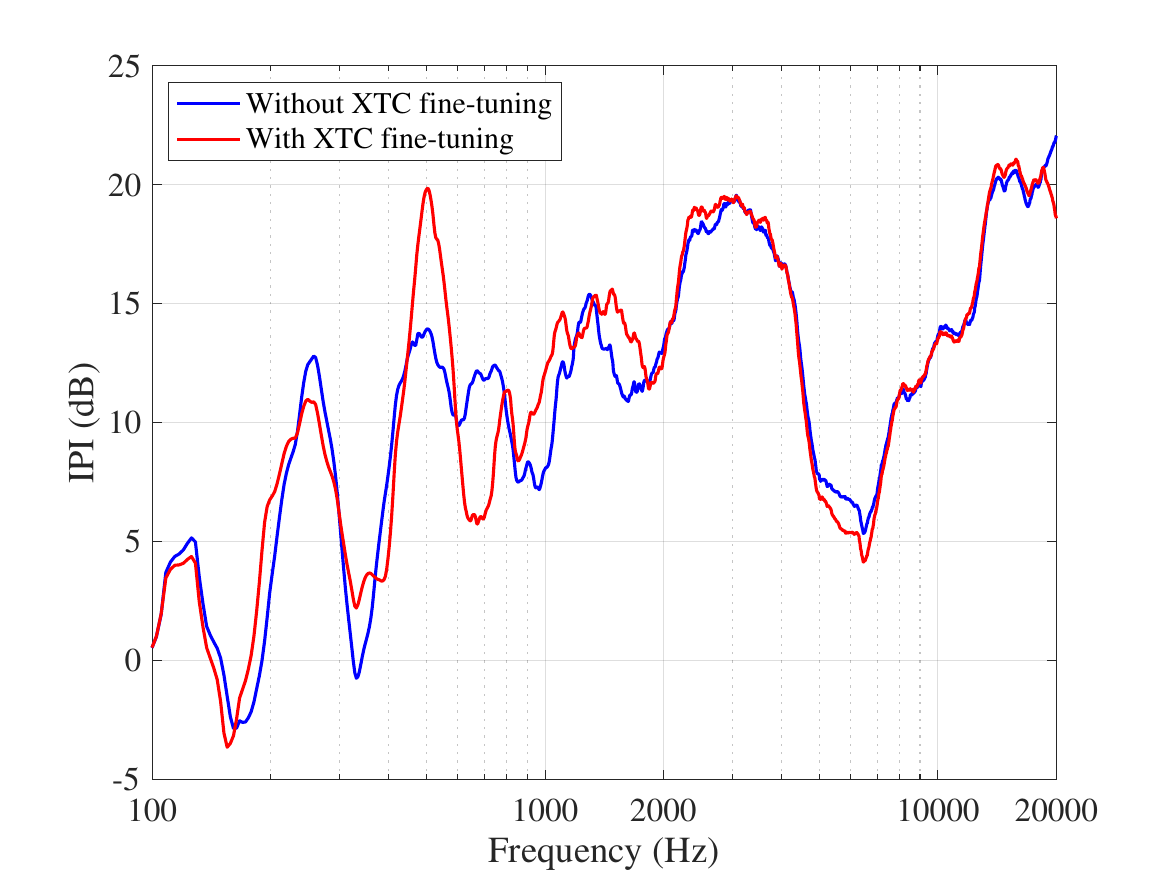}} &
\subfloat[Listener~1 XTC]{\includegraphics[width=0.32\textwidth]{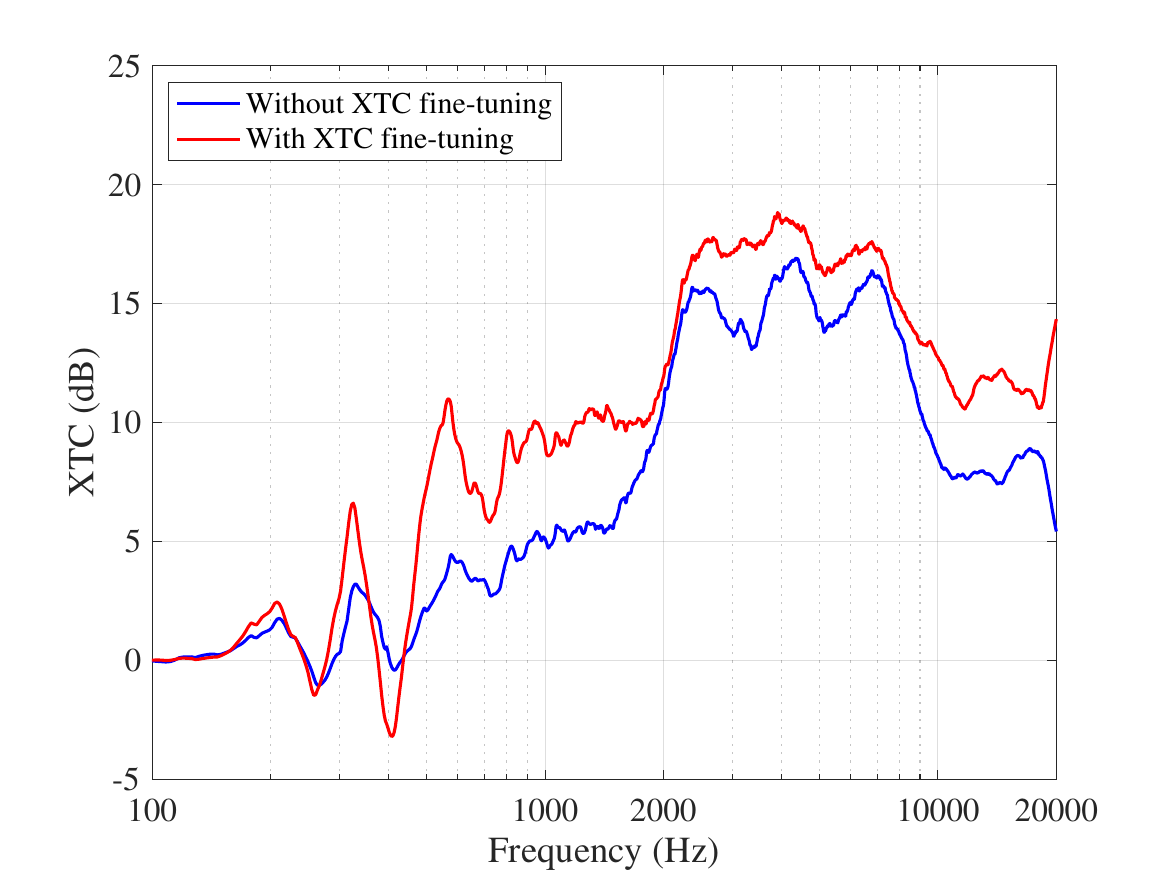}} \\[0pt]
\subfloat[Listener~2 IZI]{\includegraphics[width=0.32\textwidth]{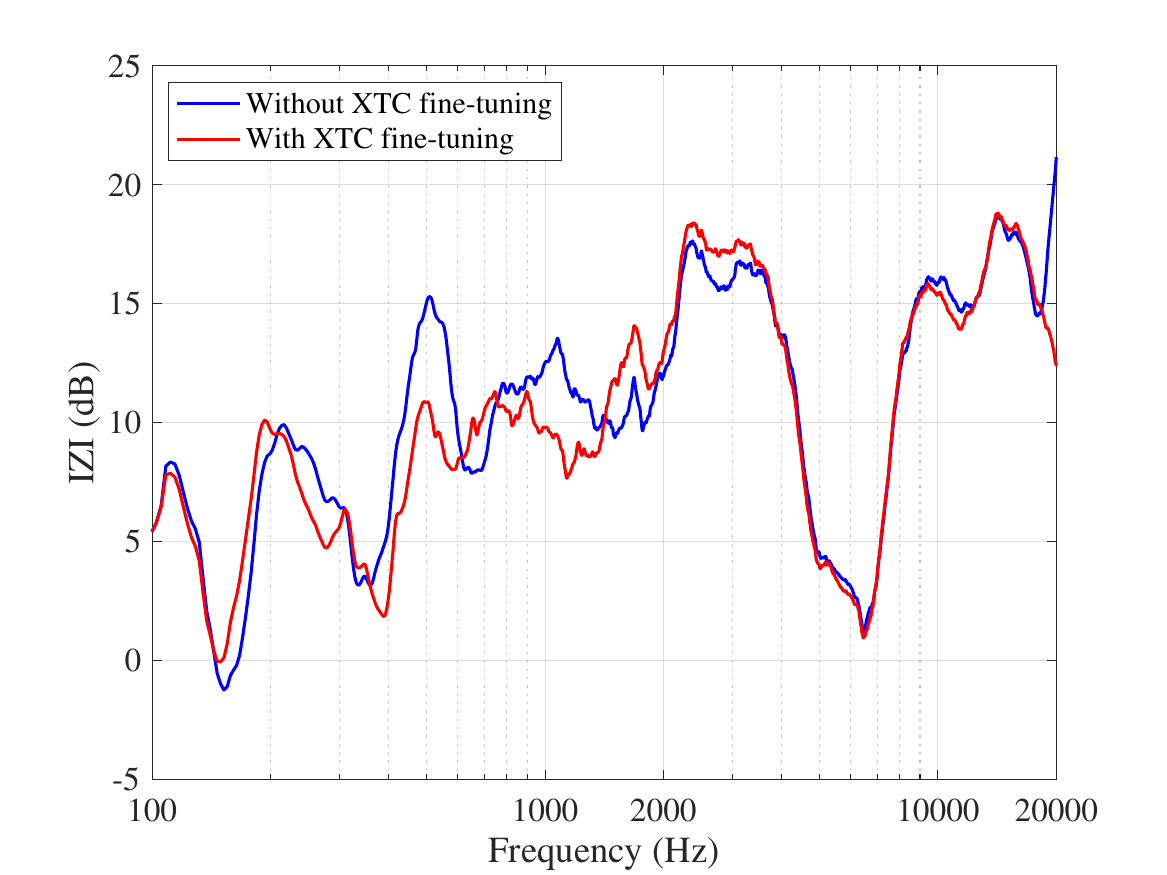}} &
\subfloat[Listener~2 IPI]{\includegraphics[width=0.32\textwidth]{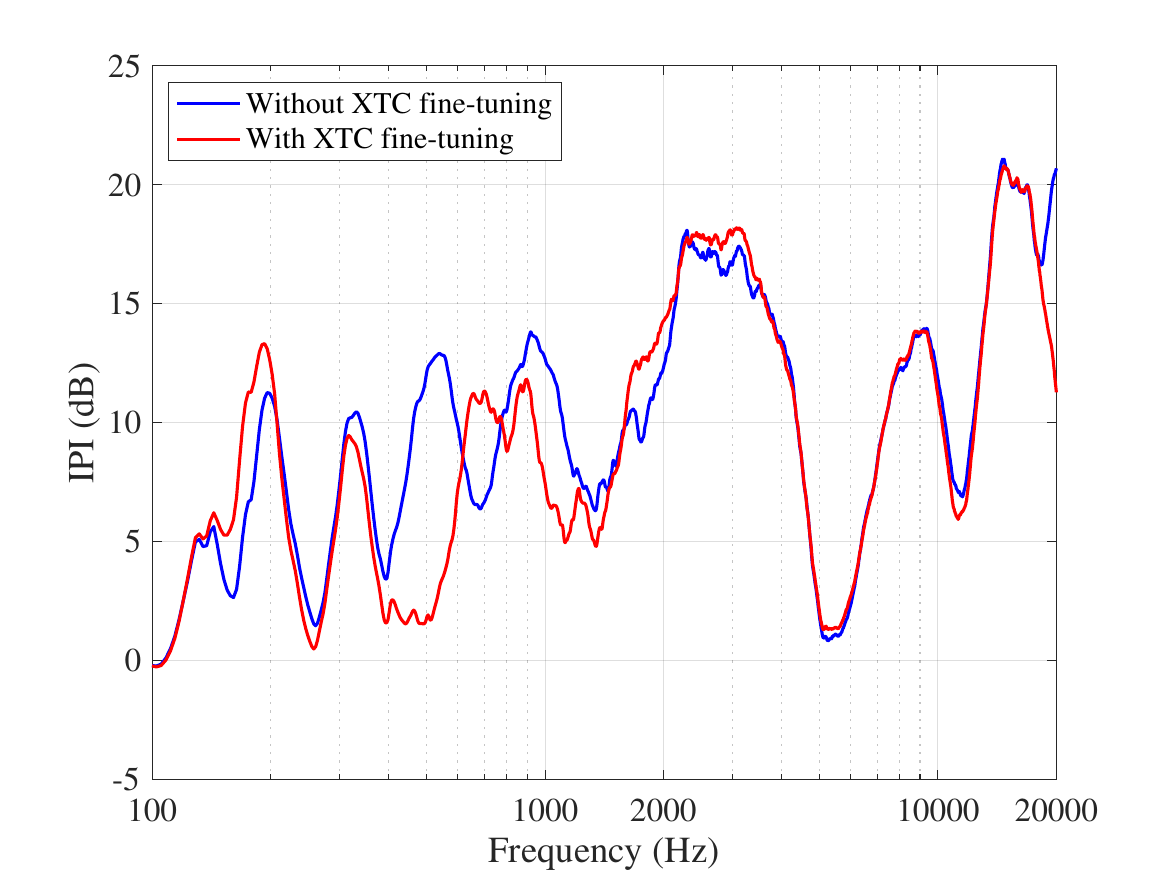}} &
\subfloat[Listener~2 XTC]{\includegraphics[width=0.32\textwidth]{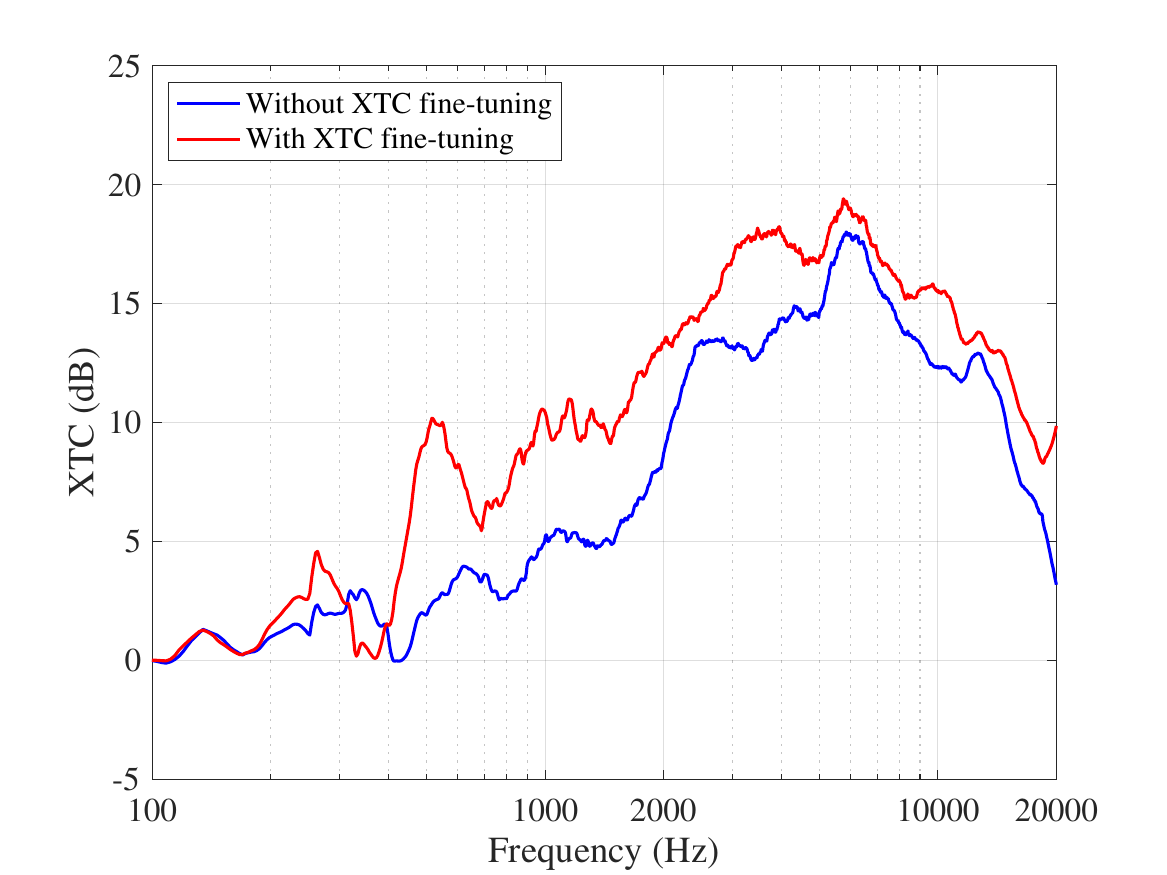}}
\end{tabular}

\caption{
Comparison between BSANN-PSZ models trained without active XTC stage (blue) and with 
active XTC stage (red), both using the same physically informed ATFs. Each panel 
shows the log-frequency--weighted IZI, IPI, or XTC curve for Listener~1 and 
Listener~2.
}
\label{fig:xtc_results}
\end{figure*}

\begin{table}[!t]
\centering
\caption{Log-frequency--weighted mean IZI, IPI, and XTC for BSANN-PSZ models 
trained with and without the active XTC stage. Both models were trained using the 
same physically informed ATFs. Subscripts 1 and 2 denote Listener~1 and 
Listener~2, respectively. The $\Delta$ column denotes the metric difference 
(Active XTC minus No XTC).\label{tab:xtc_means}}
\begin{tabular}{|c|c|c|c|}
\hline
Metric & w/o Active XTC & Active XTC & $\Delta$ \\
\hline
$\mathrm{IZI}_{1}$~[dB] & 10.23 & 10.23 & $+0.00$ \\
\hline
$\mathrm{IZI}_{2}$~[dB] & 10.43 & 10.03 & $-0.40$ \\
\hline
$\mathrm{IPI}_{1}$~[dB] & 10.87 & 11.11 & $+0.24$ \\
\hline
$\mathrm{IPI}_{2}$~[dB] & 9.79 & 9.16 & $-0.63$ \\
\hline
$\mathrm{XTC}_{1}$~[dB] & 7.93 & 10.55 & $+2.62$ \\
\hline
$\mathrm{XTC}_{2}$~[dB] & 8.19 & 11.13 & $+2.94$ \\
\hline
\end{tabular}
\end{table}

Figure~\ref{fig:xtc_results} and Table~\ref{tab:xtc_means} show the effect of active XTC stage on BSANN performance. Before applying active XTC, BSANN already 
achieves strong isolation when trained with physically informed ATFs, with 
log-frequency–weighted values of 
$\mathrm{IZI}_1 = 10.23$~dB, $\mathrm{IZI}_2 = 10.43$~dB, 
$\mathrm{IPI}_1 = 10.87$~dB, and $\mathrm{IPI}_2 = 9.79$~dB.

Applying active XTC preserves this PSZ performance. The changes in isolation 
are small, with IZI decreasing by up to 0.40~dB and IPI decreasing by up to 
0.63~dB across the two listeners, indicating that the XTC stage does not compromise the bright-zone or dark-zone control achieved during PSZ pretraining.

The primary effect of the XTC stage appears in the XTC metric. Without 
active XTC, the log-mean XTC levels are 7.93~dB (Listener~1) and 8.19~dB 
(Listener~2). With active XTC, these values increase to 10.55~dB and 11.13~dB, 
corresponding to gains of 2.62~dB and 2.94~dB, respectively. These increases 
are clearly visible in the XTC curves of Fig.~\ref{fig:xtc_results}.

Overall, active XTC stage enhances binaural XTC while maintaining the isolation performance achieved through PSZ pretraining.

\subsection{Summary}

Across all evaluation conditions, the proposed BSANN-PSZ consistently outperforms the monophonic SANN-PSZ baseline and demonstrates clear benefits from physically 
informed acoustic modeling and active XTC. Compared with SANN, BSANN provides 
substantially more balanced IZI and IPI performance across the two listeners and 
significantly improves XTC. Incorporating physically informed ATFs further 
enhances both isolation and interaural separation by capturing the loudspeaker 
frequency responses, directivity patterns, and ear-dependent scattering effects 
present in the real environment. Finally, the active XTC stage increases 
XTC levels by an additional 2.62--2.94~dB while preserving the PSZ 
performance obtained during pretraining. Together, these results provide an 
objective demonstration that stereo ear-wise control, physically informed 
modeling, and active XTC form a complementary and effective combination for 
robust multi-listener stereo and binaural PSZ rendering under realistic acoustic 
conditions.

\section{Discussion and Conclusions}
\label{sec:conclusion}

This work introduced the Binaural Spatially Adaptive Neural Network (BSANN), a unified framework for stereo and binaural PSZ rendering that optimizes the acoustic field at each ear of multiple head-tracked listeners. By extending the monophonic SANN-PSZ formulation to an ear-wise architecture, BSANN directly controls the pressures at all ears, rather than only at zone centers. Experimentally, this yields more balanced IZI and IPI across listeners in an asymmetric room and simultaneously improves XTC, indicating that the additional degrees of freedom provided by stereo ear-wise control are effectively used to counteract room-induced asymmetries and interaural leakage.

A key ingredient in this behavior is the use of physically informed ATFs that combine anechoically measured loudspeaker responses, analytic piston directivity, and rigid-sphere HRTFs. The measured loudspeaker responses embed the actual spectral coloration of the array into the training data, so the network learns filters that are naturally compatible with the deployed hardware rather than with idealized sources. The piston model introduces the frequency–angle dependence of the radiation, sharpening the main lobe at higher frequencies and shaping the side lobes, which in turn improves spatial energy steering and reduces spillover into the dark zones. The rigid-sphere HRTFs further split the field into ear-dependent components by introducing interaural level and time differences and direction-dependent scattering around the head, thereby enhancing interaural separation and stabilizing binaural cues under realistic acoustic conditions. The improvements in IZI, IPI, and passive XTC gained from replacing point-source ATFs with physically informed ATFs align with physical intuition:
the physically informed ATFs more accurately capture how the loudspeakers radiate energy in space and how each ear receives that energy, enabling the network to learn filters that better match real acoustic behavior.

The active XTC stage refines this baseline by explicitly shaping the effective ear-to-ear transfer relationships, suppressing contralateral leakage while preserving the intended ipsilateral responses. Penalizing the off-diagonal terms reduces contralateral leakage, while diagonal matching and teacher anchoring prevent the optimization from “over-correcting” and distorting the intended ipsilateral responses or the underlying PSZ field. The conditioning-aware regularizer further suppresses large filter gains in ill-conditioned frequency–position regions, which would otherwise amplify noise or room uncertainties. The measured increase of approximately 2.6–2.9~dB in XTC, with only minor changes in IZI and IPI, is in line with this interpretation.

The present study has focused on introducing the BSANN-PSZ framework and demonstrating its benefits experimentally, rather than exhaustively dissecting the contribution of each component. In particular, detailed ablation of the physically informed ATF construction (anechoically measured loudspeaker frequency responses, analytic piston directivity, and rigid-sphere HRTFs) and of the individual terms in the XTC objective is left for future work, where their relative importance and possible simplifications can be assessed systematically.

Building on the current results, several directions merit further investigation. 
First, while the BSANN framework supports head-tracked rendering in its current two-dimensional pose formulation, a natural extension is to incorporate full six-degree-of-freedom head motion by modeling both 3D translation and 3D rotation.
Second, conducting experiments across dynamic head-movement trajectories, rather than at a single evaluation pose, would more fully characterize the temporal stability and robustness of the rendered fields. 
Third, replacing the rigid-sphere model with individualized or groupwise HRTFs drawn from large datasets may further enhance binaural accuracy, especially at higher frequencies where pinna and torso effects dominate. 
Fourth, mapping the BSANN-generated frequency-domain filters to efficient IIR realizations would reduce computational cost and memory footprint for embedded or mobile deployments. Finally, although informal subjective listening tests (not reported here) suggest that most of the performance improvements described in this paper are indeed audible, systematic and exhaustive subjective evaluations will be required to fully validate the proposed BSANN-based PSZ framework.

\section*{Acknowledgment}
The authors would like to acknowledge Xiaofeng Zeng for his helpful assistance in the development of the rigid-sphere HRTF computation code used in this work.


\bibliographystyle{IEEEtran}
\bibliography{refs}

@article{druyvesteyn1997personal,
  author={W. Druyvesteyn and J. Garas},
  title={Personal sound},
  journal={J. Audio Eng. Soc.},
  volume={45},
  number={9},
  pages={685--701},
  year={1997}
}

@article{betlehem2015personal,
  author={T. Betlehem and W. Zhang and M. Poletti and T. Abhayapala},
  title={Personal sound zones: Delivering interface-free audio to multiple listeners},
  journal={IEEE Signal Process. Mag.},
  volume={32},
  number={2},
  pages={81--91},
  year={2015}
}

@article{cheer2013car,
  author  = {J. Cheer and S. J. Elliott and M. F. Sim\'{o}n G\'{a}lvez},
  title   = {Design and implementation of a car cabin personal audio system},
  journal = {J. Audio Eng. Soc.},
  volume  = {61},
  number  = {6},
  pages   = {412--424},
  year    = {2013}
}

@article{vindrola2021car,
  author={L. Vindrola and M. Melon and J.-C. Chamard and B. Gazengel},
  title={Use of the filtered-x least-mean-squares algorithm to adapt personal sound zones in a car cabin},
  journal={J. Acoust. Soc. Am.},
  volume={150},
  number={3},
  pages={1779--1793},
  year={2021}
}

@inproceedings{jacobsen2023living,
  author={R. M. Jacobsen and K. F. Skov and S. Johansen and M. B. Skov and J. Kjeldskov},
  title={Living with sound zones: A long-term field study of dynamic sound zones in a domestic context},
  booktitle = {Proc. 2023 CHI Conf. Human Factors in Computing Systems (CHI)},
  pages={1--14},
  year={2023}
}

@inproceedings{skov2023hospital,
  author={K. F. Skov and P. A. Nielsen and J. Kjeldskov},
  title={Tuning shared hospital spaces: Sound zones in healthcare},
  booktitle={Proc. 18th Int. Audio Mostly Conf.},
  pages={63--70},
  year={2023}
}

@article{heuchel2020large,
  author={F. M. Heuchel and D. Caviedes-Nozal and J. Brunskog and F. T. Agerkvist and E. Fernandez-Grande},
  title={Large-scale outdoor sound field control},
  journal={J. Acoust. Soc. Am.},
  volume={148},
  number={4},
  pages={2392--2402},
  year={2020}
}

@article{choi2002brightzone,
  author={J.-W. Choi and Y.-H. Kim},
  title={Generation of an acoustically bright zone with an illuminated region using multiple sources},
  journal={J. Acoust. Soc. Am.},
  volume={111},
  number={4},
  pages={1695--1700},
  year={2002}
}

@article{galvez2015timedomain,
  author={M. G\'{a}lvez and S. Elliott and J. Cheer},
  title={Time domain optimization of filters for personal audio},
  journal={IEEE/ACM Trans. Audio Speech Lang. Process.},
  volume={23},
  number={11},
  pages={1869--1878},
  year={2015}
}

@inproceedings{poletti2008multizone,
  author={M. Poletti},
  title={Investigation of 2D multizone surround sound systems},
  booktitle={Proc. AES Conv.},
  year={2008}
}

@article{chang2012doublelayer,
  author={J.-H. Chang and F. Jacobsen},
  title={Sound field control with a circular double-layer array of loudspeakers},
  journal={J. Acoust. Soc. Am.},
  volume={131},
  number={6},
  pages={4518--4525},
  year={2012}
}

@article{abe2023amplitude,
  author={T. Abe and S. Koyama and H. Saruwatari},
  title={Amplitude matching for multizone sound field control},
  journal={IEEE/ACM Trans. Audio Speech Lang. Process.},
  volume={31},
  pages={656--669},
  year={2023}
}

@article{lee2020signaladaptive,
  author={T. Lee and J. Nielsen and M. Christensen},
  title={Signal-adaptive and perceptually optimized sound zones with variable span trade-off filters},
  journal={IEEE/ACM Trans. Audio Speech Lang. Process.},
  volume={28},
  pages={2412--2426},
  year={2020}
}

@inproceedings{brunnstrom2022vast,
  author={J. Brunnstr{\"o}m and S. Koyama and M. Moonen},
  title={Variable span trade-off filter for sound zone control with kernel interpolation weighting},
  booktitle={Proc. ICASSP},
  year={2022}
}

@article{elliott2012robustness,
  author={S. Elliott and J. Cheer and J.-W. Choi and Y.-H. Kim},
  title={Robustness and regularization of personal audio systems},
  journal={IEEE Trans. Audio Speech Lang. Process.},
  volume={20},
  number={7},
  pages={2123--2133},
  year={2012}
}

@article{coleman2014acousticcontrast,
  author={P. Coleman and P. J. Jackson and M. Olik and M. M{\o}ller and J. Pedersen},
  title={Acoustic contrast, planarity and robustness of sound zone methods using a circular loudspeaker array},
  journal={J. Acoust. Soc. Am.},
  volume={135},
  number={4},
  pages={1929--1940},
  year={2014}
}

@article{zhu2017robustacc,
  author={Q. Zhu and P. Coleman and M. Wu and J. Yang},
  title={Robust acoustic contrast control with reduced in-situ measurement by acoustic modelling},
  journal={J. Audio Eng. Soc.},
  volume={65},
  number={6},
  pages={460--473},
  year={2017}
}

@article{moller2019tfnoise,
  author={M. B. M{\o}ller and J. K. Nielsen and E. Fernandez-Grande and S. K. Olesen},
  title={On the influence of transfer function noise on sound zone control in a room},
  journal={IEEE/ACM Trans. Audio Speech Lang. Process.},
  volume={27},
  number={9},
  pages={1405--1418},
  year={2019}
}

@inproceedings{pepe2022neuralpsz,
  author = {G. Pepe and L. Gabrielli and S. Squartini and C. Tripodi and N. Strozzi},
  title={Digital filters design for personal sound zones: A neural approach},
  booktitle = {Proc. Int. Joint Conf. Neural Networks (IJCNN)},
  pages={1--8},
  year={2022}
}

@mastersthesis{alessandri2021deep,
  author={R. Alessandri},
  title={A deep learning-based method for multi-zone sound field synthesis},
  school={Politecnico di Milano},
  year={2021}
}

@article{qiao2025sann,
  author  = {Y. Qiao and E. Y. Choueiri},
  title   = {SANN-PSZ: Spatially Adaptive Neural Network for Head-Tracked Personal Sound Zones},
  journal = {IEEE/ACM Trans. Audio Speech Lang. Process.},
  volume  = {33},
  pages   = {2735--2748},
  year    = {2025},
  doi     = {10.1109/TASLPRO.2025.3581123}
}

@article{galvez2019dynamic,
  author={M. G\'{a}lvez and D. Menzies and F. Fazi},
  title={Dynamic audio reproduction with linear loudspeaker arrays},
  journal={J. Audio Eng. Soc.},
  volume={67},
  number={4},
  pages={190--200},
  year={2019}
}

@article{ma2019superdirective,
  author={X. Ma and C. Hohnerlein and J. Ahrens},
  title={Concept and perceptual validation of listener-position adaptive superdirective crosstalk cancellation using a linear loudspeaker array},
  journal={J. Audio Eng. Soc.},
  volume={67},
  number={11},
  pages={871--881},
  year={2019}
}

@inproceedings{kabzinski2019adaptive,
  author={T. Kabzinski and P. Jax},
  title={An adaptive crosstalk cancellation system using microphones at the ears},
  booktitle={Proc. AES 147th Conv.},
  year={2019}
}

@article{lindfors2022equalization,
  author={J. Lindfors and J. Liski and V. V{\"a}lim{\"a}ki},
  title={Loudspeaker equalization for a moving listener},
  journal={J. Audio Eng. Soc.},
  volume={70},
  number={9},
  pages={722--730},
  year={2022}
}

@article{duda1998spherical,
  author={R. O. Duda and W. L. Martens},
  title={Range dependence of the response of a spherical head model},
  journal={J. Acoust. Soc. Am.},
  volume={104},
  number={5},
  pages={3048--3058},
  year={1998}
}

@article{kuhn1977spherical,
  author={G. F. Kuhn},
  title={Model for the interaural time differences in the azimuthal plane},
  journal={J. Acoust. Soc. Am.},
  volume={62},
  number={1},
  pages={157--167},
  year={1977}
}

@book{morse1968theoretical,
  author={P. M. Morse and K. U. Ingard},
  title={Theoretical Acoustics},
  publisher={Princeton Univ. Press},
  year={1968}
}

@article{Stepanishen1971Transient,
  author    = {Peter R. Stepanishen},
  title     = {Transient Radiation from Pistons in an Infinite Planar Baffle},
  journal   = {The Journal of the Acoustical Society of America},
  volume    = {49},
  number    = {5},
  pages     = {1629--1638},
  year      = {1971},
  doi       = {10.1121/1.1912548}
}

@article{qiao2022isolation,
  author={Y. Qiao and L. Guadagnin and E. Choueiri},
  title={Isolation performance metrics for personal sound zone reproduction systems},
  journal={JASA Express Lett.},
  volume={2},
  number={10},
  pages={104601},
  year={2022}
}

@inproceedings{qiao2023aes_brtf,
  author       = {Y. Qiao and E. Y. Choueiri},
  title        = {Neural Modeling and Interpolation of Binaural Room Impulse Responses with Head Tracking},
  booktitle    = {155th Conv. Audio Eng. Soc. (AES 155)},
  address      = {New York, NY},
  year         = {2023}
}

@inproceedings{tancik2020fourier,
  title={Fourier Features Let Networks Learn High Frequency Functions in Low Dimensional Domains},
  author={Tancik, Matthew and Srinivasan, Pratul P and others},
  booktitle={NeurIPS},
  year={2020}
}

@book{goodfellow2016deep,
  author    = {Goodfellow, Ian and Bengio, Yoshua and Courville, Aaron},
  title     = {Deep Learning},
  publisher = {MIT Press},
  year      = {2016}
}

@book{golub2013matrix,
  title={Matrix Computations},
  author={Golub, Gene H. and Van Loan, Charles F.},
  year={2013},
  publisher={Johns Hopkins University Press}
}

@article{DiazGuerra2021gpuRIR,
  author    = {David D{\'i}az{-}Guerra and
               Antonio Miguel and
               Juan R. Beltran},
  title     = {gpuRIR: A python library for room impulse response simulation with GPU acceleration},
  journal   = {Multimedia Tools and Applications},
  volume    = {80},
  number    = {4},
  pages     = {5653 -- 5671},
  year      = {2021},
  publisher = {Springer},
  doi       = {10.1007/s11042-020-09905-3}
  }

@incollection{choueiri2018binaural,
  author = {E. Choueiri},
  title = {Binaural audio through loudspeakers},
  booktitle = {Immersive Sound: The Art and Science of Binaural and Multi-Channel Audio},
  editor = {A. Roginska and P. Geluso},
  publisher = {Taylor and Francis},
  address = {New York, NY},
  year = {2018},
  chapter = {6},
  pages = {124--179}
}

\end{document}